\input epsf
%
%
\newfam\scrfam
\batchmode\font\tenscr=rsfs10 \errorstopmode
\ifx\tenscr\nullfont
	\message{rsfs script font not available. Replacing with calligraphic.}
\else	\font\sevenscr=rsfs7 
	\font\fivescr=rsfs5 
	\skewchar\tenscr='177 \skewchar\sevenscr='177 \skewchar\fivescr='177
	\textfont\scrfam=\tenscr \scriptfont\scrfam=\sevenscr
	\scriptscriptfont\scrfam=\fivescr
	\def\scr{\fam\scrfam}
	\def\cal{\scr}
\fi
\newfam\msbfam
\batchmode\font\twelvemsb=msbm10 scaled\magstep1 \errorstopmode
\ifx\twelvemsb\nullfont\def\Bbb{\bf}
	\message{Blackboard bold not available. Replacing with boldface.}
\else	\catcode`\@=11
	\font\tenmsb=msbm10 \font\sevenmsb=msbm7 \font\fivemsb=msbm5
	\textfont\msbfam=\tenmsb
	\scriptfont\msbfam=\sevenmsb \scriptscriptfont\msbfam=\fivemsb
	\def\Bbb{\relax\expandafter\Bbb@}
	\def\Bbb@#1{{\Bbb@@{#1}}}
	\def\Bbb@@#1{\fam\msbfam\relax#1}
	\catcode`\@=\active
\fi
%
%
\catcode`\@=11
\newfam\frakfam
\batchmode\font\twelvefrak=eufm10 scaled\magstep1\errorstopmode
\ifx\twelvefrak\nullfont\def\frak{\bf}
\message{Euler fraktur not available. Replacing with boldface.}
\else\font\tenfrak=eufm10 \font\sevenfrak=eufm7 \font\fivefrak=eufm5
\textfont\frakfam=\tenfrak
\scriptfont\frakfam=\sevenfrak \scriptscriptfont\frakfam=\fivefrak
\def\frak{\fam\frakfam}
\catcode`\@=\active\fi
%
%
\font\eightrm=cmr8	\def\xrm{\eightrm}
\font\eightbf=cmbx8	\def\xbf{\eightbf}
\font\eightit=cmti8	\def\xit{\eightit}
\font\eighttt=cmtt8	\def\xtt{\eighttt}
\font\eightcp=cmcsc8

\font\tentt=cmtt10
\font\twelverm=cmr12
\font\twelvecp=cmcsc10 scaled\magstep1

%
%
\headline={\ifnum\pageno=1\hfill\else
{\eightcp Martin Cederwall and Magnus Holm: 
	``Monopole and Dyon Spectra in N=2 SYM$\ldots$''}
		\dotfill\folio\fi}
\def\makeheadline{\vbox to 0pt{\vss\noindent\the\headline\break
\hbox to\hsize{\hfill}}
	\vskip2\baselineskip}
%
%
\footline={}
\def\makefootline{\baselineskip=1.6cm\line{\the\footline}}
%
%
\newcount\sectioncount
\sectioncount=0
\def\section#1#2{\global\eqcount=0
	\global\advance\sectioncount by 1
	\vskip2\baselineskip\noindent
	\hbox{\twelvecp\the\sectioncount. #2\hfill}\vskip\baselineskip
	\xdef#1{\the\sectioncount}}
\newcount\appendixcount
\appendixcount=0
\def\appendix#1{\global\aeqcount=0
	\global\advance\appendixcount by 1
	\vskip2\baselineskip\noindent
	\ifnum\the\appendixcount=1
	\hbox{\twelvecp Appendix A: #1\hfill}\vskip\baselineskip\fi
    \ifnum\the\appendixcount=2
	\hbox{\twelvecp Appendix B: #1\hfill}\vskip\baselineskip\fi
    \ifnum\the\appendixcount=3
	\hbox{\twelvecp Appendix B: #1\hfill}\vskip\baselineskip\fi}
%
%
\newcount\refcount
\refcount=0
\newwrite\refwrite
\def\ref#1#2{\global\advance\refcount by 1
	\xdef#1{\the\refcount}
	\ifnum\the\refcount=1
	\immediate\openout\refwrite=\jobname.refs
	\fi
	\immediate\write\refwrite
		{\item{[\the\refcount]} #2\hfill\par\vskip-2pt}}
\def\refout{\catcode`\@=11
	\xrm\immediate\closeout\refwrite
	\vskip2\baselineskip
	{\noindent\twelvecp References}\hfill
						\vskip.5\baselineskip
	\parskip=.875\parskip 
	\baselineskip=.8\baselineskip
	\input\jobname.refs 
	\parskip=8\parskip \divide\parskip by 7
	\baselineskip=1.25\baselineskip 
	\catcode`\@=\active\rm}
%
%
\newcount\eqcount
\eqcount=0
\def\Eqn#1{\global\advance\eqcount by 1
	\xdef#1{\the\sectioncount.\the\eqcount}
		\eqno(\the\sectioncount.\the\eqcount)}
\def\eqn{\global\advance\eqcount by 1
	\eqno(\the\sectioncount.\the\eqcount)}
\newcount\aeqcount
\aeqcount=0
\def\aEqn#1{\global\advance\aeqcount by 1
	\ifnum\the\appendixcount=1\xdef#1{A.\the\aeqcount}
		\eqno(A.\the\aeqcount)\fi
    \ifnum\the\appendixcount=2\xdef#1{B.\the\aeqcount}
		\eqno(B.\the\aeqcount)\fi
    \ifnum\the\appendixcount=3\xdef#1{C.\the\aeqcount}
		\eqno(C.\the\aeqcount)\fi}
\def\aeqn{\global\advance\aeqcount by 1
	\ifnum\the\appendixcount=1\eqno(A.\the\aeqcount)\fi
	\ifnum\the\appendixcount=2\eqno(B.\the\aeqcount)\fi
	\ifnum\the\appendixcount=3\eqno(C.\the\aeqcount)\fi}
	
%
%
\parskip=3.5pt plus .3pt minus .3pt
\baselineskip=12pt plus .1pt minus .05pt
\lineskip=.5pt plus .05pt minus .05pt
\lineskiplimit=.5pt
\abovedisplayskip=10pt plus 4pt minus 2pt
\belowdisplayskip=\abovedisplayskip
\hsize=15cm
\vsize=20cm
\hoffset=1cm
\voffset=1.4cm
%
%
\def\/{\over}
\def\*{\partial}
\def\a{\alpha}
\def\b{\beta}
\def\d{\delta}
\def\e{\varepsilon}
\def\g{{\frak g}}
\def\l{\lambda}
\def\ld{\lambda^\dagger}
\def\p{\psi}

\def\r{\varrho}
\def\s{\sigma}
\def\w{\omega}
\def\z{\zeta}
\def\G{\Gamma}
\def\S{\Sigma}
\def\tS{\tilde\Sigma}
\def\Z{{\Bbb Z}}
\def\C{{\Bbb C}}
\def\R{{\Bbb R}}
\def\H{{\Bbb H}}
\def\L{\Lambda}
\def\M{{\cal M}}
\def\V{{\cal V}}
\def\ad{\hbox{ad}\,}
\def\punkt{\,\,.}
\def\komma{\,\,,}
\def\.{.\hskip-1pt }
\def\is{\!=\!}
\def\-{\!-\!}
\def\+{\!+\!}
\def\={\!=\!}
\def\bra{\,<\!\!}
\def\ket{\!\!>\,}
\def\ra{\rightarrow}
\def\half{{1\/2}}
\def\quarter{{1\/4}}
\def\sdot{\!\cdot\!}
\def\cross{\!\times\!}
\def\Int{\int\!d^3x\,}
\def\Tr{\hbox{\it Tr}}
\def\f{{\lower1.5pt\hbox{$\scriptstyle f$}}}
\def\ff{{\lower1pt\hbox{$\scriptstyle f$}}}
\def\Re{\hbox{\it Re}\,}
\def\Im{\hbox{\it Im}\,}
\def\index{\hbox{\it index}}
\def\sign{\hbox{\it sign}}
\def\sti{$S^2_\infty$}
\def\Dslash{D\hskip-6.2pt/\hskip1.5pt}
\def\Lcr{\L_r^{\!\vee}}
\def\Nistwo{$N\is2$}
\def\rep{representation}
\def\csa{Cartan subalgebra}
\def\fwc{fundamental Weyl chamber}
\def\Bog{Bogomolnyi}
\def\ie{{\it i.e.}\hskip-1pt}
\def\eg{{\it e.g.}\hskip-1pt}
%
%
%
%
%

\null\vskip-1cm
\hbox to\hsize{\hfill G\"oteborg-ITP-96-4}
\hbox to\hsize{\hfill\tt hep-th/9603134}
\hbox to\hsize{\hfill March, 1996}

\epsfxsize=9cm
\vskip1.5cm\hskip2.2cm
\epsffile{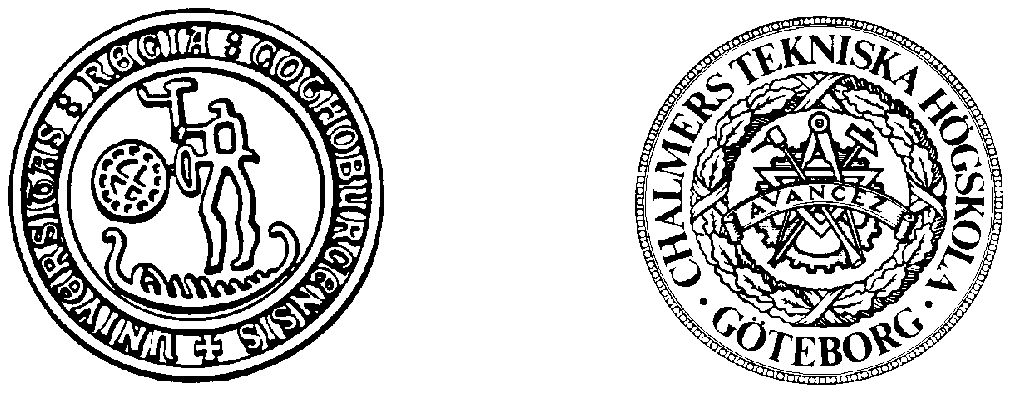}

\vskip2cm
\centerline{\twelvecp Monopole and Dyon Spectra in N=2 SYM with Higher Rank
		Gauge Groups} 
\vskip\parskip
\centerline{\twelvecp}

\vskip1.5cm
\centerline{\twelverm Martin Cederwall and Magnus Holm}

\vskip1cm
\centerline{\it Institute for Theoretical Physics}
\centerline{\it G\"oteborg University and Chalmers University of Technology }
\centerline{\it S-412 96 G\"oteborg, Sweden}

\vskip1.5cm
\noindent \underbar{\it Abstract:} We derive parts of the monopole and
dyon spectra for \Nistwo\ super-Yang--Mills theories in four dimensions
with gauge groups $G$ of rank $r\!\geq\!2$ and matter multiplets. 
Special emphasis is put on $G\is SU(3)$ and those matter contents that
yield perturbatively finite theories. There is no direct interpretation
of the soliton spectra in terms of na\"\i ve selfduality
under strong--weak coupling and exchange of electric and magnetic charges. 
We argue that, in general,
the standard procedure of finding the dyon spectrum will not give results
that support a conventional selfduality hypothesis --- the $SU(2)$ theory 
with four fundamental hypermultiplets seems to be an exception.
Possible interpretations of the results are discussed. 

\vfill
\catcode`\@=11
\hbox to\hsize{email addresses: \tentt tfemc@fy.chalmers.se\hfill}
\hbox to\hsize{\phantom{email addresses: }\tentt holm@fy.chalmers.se\hfill}
\catcode`\@=\active

\eject

\ref\SeibergIV{N.~Seiberg and E.~Witten, 
	\xit Nucl.Phys. \xbf B426 \xrm(1994) 19;
	\xrm erratum: ibid. \xbf B430 \xrm (1994) 485 ({\xtt hep-th/9407087}).}
\ref\SeibergIII{N.~Seiberg and E.~Witten, \xit Nucl.Phys. \xbf B431
	\xrm (1994) 484 ({\xtt hep-th/9408099}).}
\ref\Klemm{A.~Klemm, W.~Lerche and S.~Theisen, {\xtt hep-th/9505150}.}
\ref\Vafa{C.~Vafa and E.~Witten, \xit Nucl.Phys. \xbf B431 \xrm (1994) 3
	({\xtt hep-th/9408074}).}
\ref\IntriligatorI{K.~Intriligator and N.~Seiberg \xit Nucl.Phys. \xbf B431 
	\xrm (1994) 551 ({\xtt hep-th/9408155}).}
\ref\Montonen{C.~Montonen and D.~Olive, 
	\xit Phys.Lett. \xbf 72B \xrm (1977) 117.}
\ref\GNO{P.~Goddard, J.~Nuyts and D.~Olive, 
	\xit Nucl.Phys. \xbf B125 \xrm (1977) 1.}
\ref\Mandelstam{S.~Mandelstam, \xit Nucl.Phys. \xbf B213 \xrm (1983) 149.}
\ref\Brink{L.~Brink, O.~Lindgren and B.E.W.~Nilsson,
	\xit Phys.Lett. \xbf 123B \xrm (1983) 323.}
\ref\Sohnius{M.~Sohnius and P.~West, \xit Phys.Lett. \xbf 100B \xrm (1981) 45.}
\ref\Howe{P.S.~Howe, K.S.~Stelle and P.C.~West, \xit Phys.Lett. \xbf 124B
	\xrm (1983) 55.}
\ref\HoweII{P.S.~Howe, K.S.~Stelle and P.K.~Townsend,
	\xit Nucl.Phys. \xbf B214 \xrm (1983) 519.}
\ref\Piguet{O.~Piguet and K.~Sibold, 
	\xit Helv.Phys.Acta \xbf 63 \xrm (1990) 71.}
\ref\SeibergV{N.~Seiberg \xit Phys.Lett. \xbf 206B \xrm (1988) 75.}
\ref\Sen{A.~Sen, \xit Phys.Lett. \xbf B329 \xrm (1994) 217
	({\xtt hep-th/9402032}).}
\ref\Porrati{M.~Porrati, {\xtt hep-th/9505187}.}
\ref\GauntlettIV{J.P.~Gauntlett and D.A.~Lowe, {\xtt hep-th/9601085}.}
\ref\Lee{K.~Lee, E.~Weinberg and P.~Yi, {\xtt hep-th/9601097}.}
\ref\Girardello{L.~Girardello, A.~Giveon, M.~Porrati and A.~Zaffaroni, 
      \xit Nucl.Phys. \xbf B448 \xrm (1995) 127 ({\xtt hep-th/9502057}).}
\ref\Bogomolnyi{E.B.~Bogomolnyi, 
	\xit Sov.J.Nucl.Phys. \xbf 24 \xrm (1976) 449.}
\ref\Prasad{M.K.~Prasad and C.M.~Sommerfield, \xit Phys.Rev.Lett. \xbf 35
	\xrm (1975) 760.}
\ref\Cederwall{M.~Cederwall, G.~Ferretti, B.E.W.~Nilsson and P.~Salomonson,
	{\xtt hep-th/9508124}, \xit Mod.Phys.Lett. \xrm in press.} 
\ref\Harvey{J.A.~Harvey and A.~Strominger, \xit Commun.Math.Phys. \xbf
	151 \xrm (1993) 221 ({\xtt hep-th/9108020}).}
\ref\Callias{C.~Callias, \xit Commun.Math.Phys. \xbf 62 \xrm (1978) 213.}
\ref\Weinberg{E.~Weinberg, \xit Nucl.Phys. \xbf B167 \xrm (1980) 500;
	\xit Nucl.Phys. \xbf B203 \xrm (1982) 445.}
\ref\Hooft{G.~`t~Hooft, \xit Nucl.Phys. \xbf B79 \xrm (1976) 276.}
\ref\Polyakov{A.M.~Polyakov, \xit JETP Lett. \xbf 20 \xrm (1974) 194.}
\ref\Sethi{S.~Sethi, M.~Stern and E.~Zaslow, {\xtt hep-th/9508117}.}
\ref\GauntlettIII{J.P.~Gauntlett and J.A.~Harvey, {\xtt hep-th/9508156}.}
\ref\Barbon{J.L.F.~Barb\'on and S.~Ramgoolam, {\xtt hep-th/9512063}.}
\ref\Olive{D.~Olive and E.~Witten, \xit Phys.Lett. \xbf 78B \xrm (1978) 97.}
\ref\MantonII{N.S.~Manton and B.J.~Schroers, \xit Ann.Phys. \xbf 225 
	\xrm (1993) 290.}
\ref\Pope{C.~Pope, \xit Nucl.Phys. \xbf B141 \xrm (1978) 432;
	\xit J.Phys. \xbf A14 \xrm (1981) L133.}
\ref\Hanany{A.~Hanany and Y.~Oz, {\xtt hep-th/9505075}.} 
\ref\ArgyresI{P.C.~Argyres, M.R.~Plesser and A.D.~Shapere,
	{\xtt hep-th/9505100}.}
\ref\ArgyresII{P.C.~Argyres, M.R.~Plesser, and N.~Seiberg,
	{\xtt hep-th/9603042}.}
\ref\Minahan{J.A.~Minahan and D.~Nemeschansky, {\xtt hep-th/9507032};
	{\xtt hep-th/9601059}.}
\ref\ArgyresIII{P.C.~Argyres and M.R.~Douglas, 
      \xit Nucl.Phys. \xbf B448 \xrm (1995) 93 ({\xtt hep-th/9505062}).} 
\ref\Aharony{O.~Aharony and S.~Yankielowicz, {\xtt hep-th/9601011}.}	
\ref\AtiyahI{M.F.~Atiyah and N.~Hitchin, \xit ``The Geometry and Dynamics of 
	Magnetic Monopoles'',\hfill\break\indent
	 \xrm Princeton University Press, Princeton (NJ) (1988).}
\ref\Donaldson{S.K.~Donaldson, \xit Commun.Math.Phys. \xbf 96 \xrm (1984) 387.}
\ref\Hurtubise{J.~Hurtubise, \xit Commun.Math.Phys. \xbf 120 \xrm (1989) 613.}
\ref\Gibbons{G.W.~Gibbons and P.J.~Ruback, 
	\xit Commun.Math.Phys. \xbf 115 \xrm (1988) 267.}
\ref\AtiyahIII{M.F.~Atiyah, V.K.~Patodi and I.M.~Singer,
	\xit Math.Proc.Camb.Phil.Soc. \xbf 77 \xrm (1975) 43.}

\section\introduction{Introduction}
The last years have seen a tremendous progress in the understanding
of nonperturbative aspects of four-dimensional field theory.
New techniques [\SeibergIV,\SeibergIII,\Klemm,\Vafa,\IntriligatorI] 
enable calculation of exact results valid beyond
the perturbative level. It was long ago conjectured [\Montonen,\GNO] that
the $N\is4$ supersymmetric Yang--Mills (SYM) theories should possess
some kind of strong--weak coupling duality. These theories are 
perturbatively finite [\Mandelstam,\Brink,\Sohnius,\Howe,\HoweII,\Piguet], 
and actually exactly finite [\SeibergV].
Actual calculations of dyon spectra in these theories [\Sen,\Porrati,
\GauntlettIV,\Lee], and also other tests [\Girardello] give
strong support for the duality hypothesis.
There are also an infinite number of theories, possessing $N\is2$, but
not $N\is4$, supersymmetry that are one-loop, and thus perturbatively,
finite. The only one of these theories that has undergone closer
examination with respect to duality properties is the $SU(2)$ model
with four hypermultiplets in the fundamental \rep. There, all results
confirm duality, and it is tempting to conclude that the same is
true for all perturbatively finite $N\is2$ SYM theories.
Since all explicit calculations of BPS states in $N\is4$ theories and the
finite $N\is2$ $SU(2)$ theory sofar are in excellent agreement with
predictions from duality, it is natural to continue this program and
include also the other perturbatively finite $N\is2$ theories.
The aim of this paper is to do this by calculating part of the dyon spectra
for such theories. As we will demonstrate, a number of problems arise.
They are partly associated with the lattice structures of electric and magnetic
charges, and also with the inaccessibility of monopole--anti-monopole
configurations.

In sections 2 and 3, basic properties about monopoles
and their moduli spaces are reviewed. Section 4 applies an index theorem
to find the dimensions of bundles of zero-modes of the various fields 
in the theories over moduli space. Section 5 contains a discussion on
the lattice properties of electric and magnetic charges, giving a
general argument against na\"\i ve duality. In section 6, the effective
action for the monopoles is derived from the field theory, and some aspects
of its quantization are discussed. Section 7 applies this quantization
to some specific examples, and derives the corresponding dyon spectra.
They do not support na\"\i ve duality.
In section 8, the implications of the results are discussed.

\section\monopoles{Monopoles --- Symmetry Breaking and Topology}
In this section, we will give a quick review of the concept of 
\Bog--Prasad--Sommerfield (BPS) monopoles [\Bogomolnyi,\Prasad]
and their topological properties, aiming at a topological description
suited for the index calculations of section 4. 
A BPS (multi-)monopole is a static
configuration of the Yang--Mills--Higgs (YMH) system that due to 
its topological
character has a relation between mass (energy) and magnetic charge.
Consider the hamiltonian of the YMH system with gauge group $G$
(the Higgs field is in the adjoint \rep):
$$
H=\half\int d^3x\Tr\left(B_iB_i+D_i\Phi D_i\Phi\right)
=\quarter\int d^3x\Tr\left\{\left(B_i+D_i\Phi\right)^2
+\left(B_i-D_i\Phi\right)^2\right\}\punkt\eqn 
$$
If the \Bog\ equation 
$$
B_i=\pm D_i\Phi\Eqn\bogo
$$
is imposed (note that this equation alone implies that the equations
of motion are satisfied), the energy becomes topological:
$$
H=\pm\int d^3x\Tr\,B_iD_i\Phi=\int_{\R^3}\Tr\,FD\Phi
=\int_{\R^3}\Tr\,D(F\Phi)=\int_{S^2_\infty}\Tr\,F\Phi\komma\Eqn\topologicalmass
$$
and can be related to the topological magnetic charges of the field
configuration (see below).

The topological information of the BPS configuration resides entirely in the
asymptotic behaviour of the Higgs field. Let us denote the Higgs field at
the two-sphere \sti at spatial infinity by $\phi(x)$. 
By a gauge transformation,
it can always (locally) be brought to an element in the \csa\ (CSA) of $\g$,
the Lie algebra of $G$, and furthermore, by Weyl reflections, into a \fwc.
The equations of motion then imply that this element is constant on \sti.
We thus have 
$$
\psi=g^{-1}(x)\phi(x) g(x)\komma\Eqn\psiphi
$$
where $\psi$ is a constant element in the CSA. The group element $g(x)$ is
not globally defined on \sti, though $\phi$ and $\psi$ are. If $g$ is 
defined patchwise on the two hemispheres, the difference on the equator is
an element in $H\subset G$, the stability group of $\phi$. 
We will only consider the generic case of maximal symmetry breaking,
when $H$ is the maximal torus of $G$. This occurs as long as the 
diagonalized Higgs field $\psi$ does not happen to be orthogonal to
any of the roots. $H\=(U(1))^r$ is the unbroken gauge group,
where $r$ is the rank of $G$.
In the light of equation (\psiphi), the Higgs field on \sti may be viewed
as a map from \sti to the homogeneous space $G/H$, and all the topological
information now lies in the gauge transformation $g(x)$. The relevant
classification is $\pi_2(G/H)$, which (for semisimple $G$) is isomorphic
to $\pi_1(H)$. For the case at hand, this group is $\Z^r$, \ie\ there are
$r$ magnetic charges. 
It is straightforward to calculate the vector $k$ of
magnetic charges. The gauge transformation (\psiphi) induces a
connection 
$\w\=g^{-1}dg$ with field strength
$f\=d\w\+\w^2\=0$ locally but not globally (with the two patches defined above,
$f$ has distributional support on the equator), the magnetic charges of which
can be expressed as
$$
k\sdot T={1\/2\pi i}\int_{S^2}f
	={1\/2\pi i}\int_{S^1}(\w_{\hbox{\xrm north}}
	-\w_{\hbox{\xrm south}})
		\Eqn\chargevector
$$
(the last integral is evaluated at the equator of $S^2$ where the two patches
of the connection meet).
The mass of the configuration is expressed in terms of $k$ 
using equation (\topologicalmass):
$$
m=\pm\int_{S^2}\Tr\,f\p=2\pi\,|h\sdot k|\Eqn\masschargerelation
$$
where $\psi$ is expressed in terms of the vector $h$ as 
$\psi\=h\sdot T\in\hbox{CSA}$. In section 4, we will use the gauge 
transformation $g$ in order to calculate indices of Dirac operators
in a monopole background, yielding the number of zero-modes of certain
fields in the presence of a monopole.

The magnetic charge vector obtained from equation (\chargevector) lies
on the coroot lattice $\Lcr$ of $G$. This agrees with the generalized 
Dirac quantization condition on electric and magnetic charges, that
$$
e\sdot k\in\Z\Eqn\diraccondition
$$ 
for any charge vectors $e$ and $k$. Since $e$ must
lie on the weight lattice $\L_w$ of $G$, $k$ must lie on the dual lattice
of the weight lattice, \ie\ the coroot lattice. We should comment on our
choice of normalization for the magnetic charges. It means that the scale
of the coroot lattice is chosen so that the coroots are 
$$
\Lcr\ni\a^{\!\vee}={2\a\/\,|\a|^2}\komma\Eqn\corootnormalization
$$ 
and coincide with the roots for simply laced groups.

An elegant and convenient way of treating the YMH system in a unified way
is to consider the Higgs field as the fourth component of a euclidean
four-dimensional gauge connection. We thus let $A_4\=\Phi$, and demand that
no fields depend on $x^4$. It is useful to go to a quaternionic formalism,
where the gauge connection sits in a quaternion $A\=A_\mu e_\mu\in\H$, 
$e_4\is1$ being
the quaternionic unit element and $e_i$, $i\is1,2,3$  the imaginary
unit quaternions: $e_ie_j\=-\delta_{ij}\+\e_{ijk}e_k$.
The \Bog\ equation (\bogo) now becomes an (anti-)selfduality equation
for the field strength $F_{\mu\nu}$:
$$
F_{\mu\nu}=\pm\half\e_{\mu\nu\rho\sigma}F_{\rho\sigma}\Eqn\selfdualF
$$
and the topological character of the solutions becomes even more obvious.
We will use the fact that a selfdual antisymmetric tensor can be expressed
as an imaginary quaternion, and is formed from two vectors as
$f^+\=\Im(vw^*)$. An anti-selfdual tensor is formed as $f^-\=\Im(v^*w)$.
Spinors of both chiralities come as quaternions. The Weyl equations
are for the $s$ chirality $D^*s\is0$ and for the $c$ chirality $Dc\is0$.
For a more detailed discussion of the quaternionic formalism, 
transformation properties etc., see \eg\
reference [\Cederwall].

\section\zeromodes{Moduli Spaces and Zero-Modes}
A monopole solution is not an isolated phenomenon. There are always 
deformations of the field configuration that do not modify the energy.
These always continue to satisfy the \Bog\ equation (\bogo, \selfdualF).
Deformations of the YMH system alone define tangent directions in the
moduli space of monopole solutions at given magnetic charge $k$.
One obvious set of such deformations is given by simply translating
the (localized) solution. Therefore, the moduli space always contains
a factor $\R^3$, but there are in general more possible moduli.
Also, when other fields are present, as in the \Nistwo\ models we consider,
these may also possess zero-modes in the BPS monopole background.
These zero-modes also have to be considered in the low energy treatment
we will make.

We will first give a resum\'e of some of the geometric aspects of
the geometry of the moduli spaces (following reference [\Harvey], but
in the quaternionic formalism of [\Cederwall]), 
and then move on to the full
\Nistwo\ model.

Suppose we search for a deformation $\d A$ of the gauge connection
(in a quaternionic form, containing the Higgs field). The linearized 
version of the \Bog\ equation (with the plus sign --- the anti-selfdual
case is analogous) is $\Im(D^*\d A)\=0$,
where the rule for formation of an anti-selfdual tensor from from two
vectors has been used. Denote the tangent directions by an index $m$.
The natural metric is induced from the kinetic term in the action,
$$
g_{mn}=\Int\Tr(\d_m A_\mu\d_n A_\mu)=\Int\Tr\Re(\d_m A^*\d_n A)
	\equiv\bra\d_m A,\d_n A\ket\punkt
\Eqn\modulimetric
$$
We would like a (physical) tangent vector to be orthogonal to any
gauge modes in this metric, and therefore impose the supplementary
condition $\Re(D^*\d A)\=0$. The two conditions so derived for the
deformations $\d_mA$ are collected in
$$
D^*\d_mA=0\punkt\Eqn\tangenteq
$$
We note that this equation is formally identical to a Weyl equation
for one of the four-dimensional spinor chiralities.
It is also straightforward to show that the Weyl equation for the other
chirality never can have $L^2$ solutions, simply because the background
field strength is selfdual. The dimension of a moduli space at given $k$
can therefore be calculated as the $L^2$ index of the Dirac operator on
$\R^3$ in a known BPS background. As we will see, the only essential
information that goes into the index calculation is the asymptotic 
behaviour of the Higgs field. This calculation will yield the complex
dimension of the moduli space, {\it provided some selfdual solution with this
asymptotic behaviour exists}.

All moduli spaces are known to be hyperK\"ahler. The action of the complex
structures on the tangent vectors is easily understood. If a tangent
vector $\d_mA$ satisfies equation (\tangenteq), then also 
$\d_mAe_i$ satisfy the same equation. The three complex structures act as
$$
J^{(i)\phantom{m}n}_{\phantom{(i)}\raise4pt\hbox{$\scriptstyle m$}}
	\d_n A=\d_m Ae_i\punkt\Eqn\complexstructures
$$
They can be shown to be covariantly constant with respect to the 
connection derived from (\modulimetric). 

A parallel transport in the tangent directions of moduli space on the space 
of zero-modes should preserve the condition that tangent vectors are 
orthogonal to gauge modes. In order to achieve this, one introduces
the gauge parameters $\e_m(x)$ and writes
$$
\d_m A=\*_m A-D\e_m\punkt\eqn
$$
Parallel transport is generated by the covariant derivative
$s_m\is\*_m\+\ad\e_m$ (more generally, $\e_m$ acts in the appropriate
\rep\ of the gauge group), with the property $[s_m,D]\=\d_m A$.
This implies that $D^*_{A+dt^m\d_mA}(\r\+dt^ms_m\r)\=0$ for zero-modes
in any \rep, so that $s_m$ provides a good parallel transport of all
zero-modes. It is straightforward to calculate the Christoffel connection
of the metric (\modulimetric),
$$
{\G^m}_{np}=g^{mq}\Int\Tr\,\,\d_q A_\mu s_n\d_p A_\mu
	=g^{mq}\Int\Tr\,\Re(\d_q A^*s_n\d_p A)
	=g^{mq}\!\bra\d_q A,s_n\d_p A\ket\komma
	\eqn
$$
and the riemannian curvature [\Cederwall],
$$
\eqalign{
R_{mnpq}&=\bra\d_p A,[s_m,s_n]\d_q A\ket+\bra s_m\d_p A,\Pi_+s_n\d_q A\ket
				-\bra s_n\d_p A,\Pi_+s_m\d_q A\ket\cr
	&=\bra\d_p A,[s_m,s_n]\d_q A\ket
	-4{P_{+pq}}^{rs}\!\bra\d_mA,[s_n,s_r]\d_sA\ket\komma\cr}
	\Eqn\curvature
$$
where $\Pi_+\is D(D^*\!D)^{-1}\!D^*$ 
is the projection operator on higher modes and 
${P_{+pq}}^{rs}\={1\/4}{{J^{(a)}}_{[p}}^{\lower2pt\hbox{$\scriptstyle r$}}
{{J^{(a)}}_{q]}}^{\lower2pt\hbox{$\scriptstyle s$}}$ is the projection 
operator on the part of
an antisymmetric tensor that commutes with the complex structures,
i.e\.\ the $Sp(n)$ part, $4n$ being the real dimension 
($J^{(4)}$ is defined as the unit matrix). The curvature is a $(1,1)$-form
with respect to all three complex structures, which is equivalent to
$Sp(n)$ holonomy, \ie\ ``selfduality''.
     
The action for our \Nistwo\ super-Yang--Mills theory with matter is
most conveniently formulated as the dimensional reduction of an $N\is1$
theory in $D\is(1,5)$. The six-dimensional action reads:
$$
\eqalign{{\cal L}=-{1\/4}F_{MN}&F^{MN}+\half\Re(\ld\S^MD_M\l)\cr-
		&\half D_Mq^*_\ff D^Mq_\f
	+\half\Re(\p^\dagger_f\tS^MD_M\p_\f)+\Re(\p^\dagger_f\l q^*_\ff)
	+{1\/8}(q^*_\ff\cross q_\f)^2\punkt\cr}
	\Eqn\fieldtheoryaction
$$
Here, \rep\ indices and traces have been suppressed for clarity.
In addition to the gauge potential and its superpartner $\l$ in the adjoint
\rep, there are the matter bosons $q$ and fermions $\p$. The subscript
$f$ labels the matter multiplets. A dagger denotes quaternionic conjugation
and transposition, and, if the \rep s of $G$ are complex, 
also complex conjugation. The matrices $\S$ and $\tS$ are six-dimensional
quaternionic sigma matrices, and the cross product in the last term
denotes Clebsh--Gordan coefficients for formation of an element in the
adjoint \rep. The fermions $\l$ and $\p$ are two-component quaternionic
spinors of opposite six-dimensional chiralities, and the matter boson $q$
is a scalar quaternion.

The supersymmetry transformations are:
$$
\matrix{\d A_M=\Re(\e^\dagger\S_M\l)\komma\hfill
		&\d q_\f=\p^\dagger_f\e\komma\hfill\cr
	\lower4pt\hbox{$\d\l=-\half F_{MN}\tS^{MN}\e
		+\half\e(q^*_\ff\cross q_\f)\komma$}\hfill
	&\lower8pt\hbox{$\d\p_\f=\S^M\e D_Mq^*_\ff\punkt$}\hfill\cr}
	\Eqn\supersymmetrytransformations
$$
It is clear from the transformation of $\l$ that a BPS background, obeying
(\bogo), breaks half the supersymmetry.

The Higgs field comes as one of the components ($A_4$, say) of the 
six-dimensional gauge connection. The euclidean four-dimensional
formulation automatically comes out on reduction to four euclidean
dimensions, upon which a spinor (of any six-dimensional chirality) splits into
a pair of quaternionic spinors of opposite four-dimensional chiralities.  

In order to examine which fields carry zero-modes in the BPS background,
and go into a low energy expansion, we give the moduli parameters a slow
time dependence
and expand the equations of motion in the parameter 
$n\is \#({d\/dt})\+\half\#(\hbox{fermions})$. At $n\is0$ one only has
the background fields $A$ with selfdual field strength. At $n\is\half$,   
there are the Weyl equations for the upper ($s$ chirality) components
of $\l$ and $\p$, which we denote $\a$ and $\b$, respectively. 
Their lower ($c$ chirality) components vanish to this order.
The time dependence of the bosonic moduli is modeled so that
$A\=A(x,X(t))$. Then the equations at order $n\is1$ imply,
using $\dot A\is\dot X^m(\d_m A\+D\e_m)$,
$$
\eqalign{
&A_0=\dot X^m\e_m+(D^*\!D)^{-1}(-\a^*\a+\half\b_\ff^*\cross\b_\f)\komma\cr
&A_5=(D^*\!D)^{-1}(\a^*\a+\half\b_\ff^*\cross\b_\f)\komma\cr
&q^*_\ff=-(D^*\!D)^{-1}(\a^*\b_\f)\punkt\phantom{\half}\cr}\Eqn\nisonesolution
$$
We see that the only fields that carry zero-modes, apart from the tangent
directions to moduli space itself in the YMH system, are the fermions,
both in the vector multiplet and the matter multiplets.

In order to get information about the number of fermionic zero-modes
in the BPS background, we have to apply the index theorem of Callias
[\Callias] to the appropriate \rep s of $\l$ (the adjoint) and $\p$.
We have already seen that the equation for tangent vectors to the moduli
space is equivalent to a Weyl equation, so that the zero-modes of
$\l$ will come in the tangent bundle over moduli space, whose dimension
is given by the index theorem. The zero-modes of $\p$ will come in
some other index bundles with some connections. These connections and
their curvatures are derived analogously to the riemannian curvature above.
if the mode functions are denoted $\r_\a$ and normalized so that $\a$
is the fiber index of an orthonormal bundle, the connection is
$$
\w_{m\a\b}=\bra\r_\a,s_m\r_\b\ket\komma\eqn
$$
and the curvature
$$
F_{mn\a\b}=\bra\r_a,[s_m,s_n]\r_\b\ket
	+\bra s_m\r_\a,\Pi_+s_n\r_\b\ket
	-\bra s_n\r_\a,\Pi_+s_m\r_\b\ket\punkt\Eqn\indexcurvature
$$

\section\indexcalculations{Dimensions of Moduli Spaces and Index Bundles}
Callias [\Callias] has given an index theorem for the Dirac operator
on $\R^{2n\-1}$ in the presence of a gauge connection and a scalar
matrix valued hermitean (Higgs) field that takes some nonzero values 
at spatial infinity.
This index theorem is applicable precisely to the situation at hand.
The index only depends on the (topological) behaviour of the Higgs
field $\Phi$ at infinity. Callias theorem states that the $L^2$ index 
of the Dirac operator on $\R^3$ in the representation $\r$ is given as
$$
\index\Dslash_\r=-{1\/16\pi i}\int_{S^2_\infty}\Tr_\r(UdUdU)\punkt
	\Eqn\calliastheorem
$$
Here, the matrix $U$ is defined as $U\is(\phi^2)^{-1/2}\phi$. Callias
postulates that $\phi$ should have no zero eigenvalues, so that $U$
is well defined. This assumption is directly related to the Dirac
operator being Fredholm. If it does not have this property, there is
a continuous spectrum around zero that, depending on the behaviour
of the density of states, may contribute to the index calculation
and give an incorrect result. Actually, in the case we are interested
in, there are zero eigenvalues, corresponding to the fields that remain
massless after the symmetry breaking. E.~Weinberg [\Weinberg] has shown that
the massless vector bosons of the generic maximal symmetry breaking
pattern do not contribute in the index calculation. On the other hand,
for nonmaximal breaking to a nonabelian group $H$, one has to be more
careful, and examine the exact contribution due to the roots orthogonal
to the Higgs field. The same is true for some special values of the Higgs
field that becomes orthogonal to some weight in a representation
for the matter fields (see below). In the generic case, though, all one
has to do is to replace the matrix $\phi$ by its restriction to the
subspace spanned by the eigenvectors with nonzero eigenvalues. The 
corresponding restricted Dirac operator will have the 
desired Fredholm property.

The actual computation of the index is conveniently performed using
the gauge transformation $g$ of section \monopoles. After the gauge 
transformation has been performed, the Higgs field has changed to
the diagonalized Higgs field $\p\in\hbox{CSA}$, and the derivative simply 
becomes the commutator with the induced connection $\w$, since $d\p\is0$.
We thus have
$$
\index\Dslash_\r=-{1\/16\pi i}\int_{S^2_\infty}\Tr_\r(V[\w,V]^2)
	={1\/4\pi i}\int_{S^2_\infty}\Tr_\r(Vd\w)\komma
	\Eqn\calliastheoremii
$$
where $V\is(\psi^2)^{-1/2}\psi$, and we have used $V^2\is1$ and 
$d\w\is-\w^2$. Taking the trace in the representation $\r$ gives the
result, using equation (\chargevector) for the magnetic charge vector,
$$
\eqalign{
&\index\Dslash_\r=k\sdot\L\komma\cr
&\L=\half\sum_{\l\in\r}\l\,\sign(h\sdot\l)\komma\cr}\Eqn\explicitindex
$$
the sum being performed over the weights of the representation $\r$.
It is clear that the index stays constant as long as $h$ does not 
become orthogonal to some weight, in which case the index changes 
discontinuously. 

The expression (\explicitindex) enables us to calculate the index
explicitly for any magnetic charge and any \rep\ of $G$. We will now turn 
to some examples that will be of use later.
We first define the simple roots with respect to the 
value of the diagonalized Higgs field in the \csa. The vector $h$ can
always be chosen in the \fwc\ so that its scalar product with all simple
roots is positive.
This is illustrated for $SU(3)$ in figure 1.

\epsfxsize=.4\hsize
\vskip.4cm
\hskip3cm\epsffile{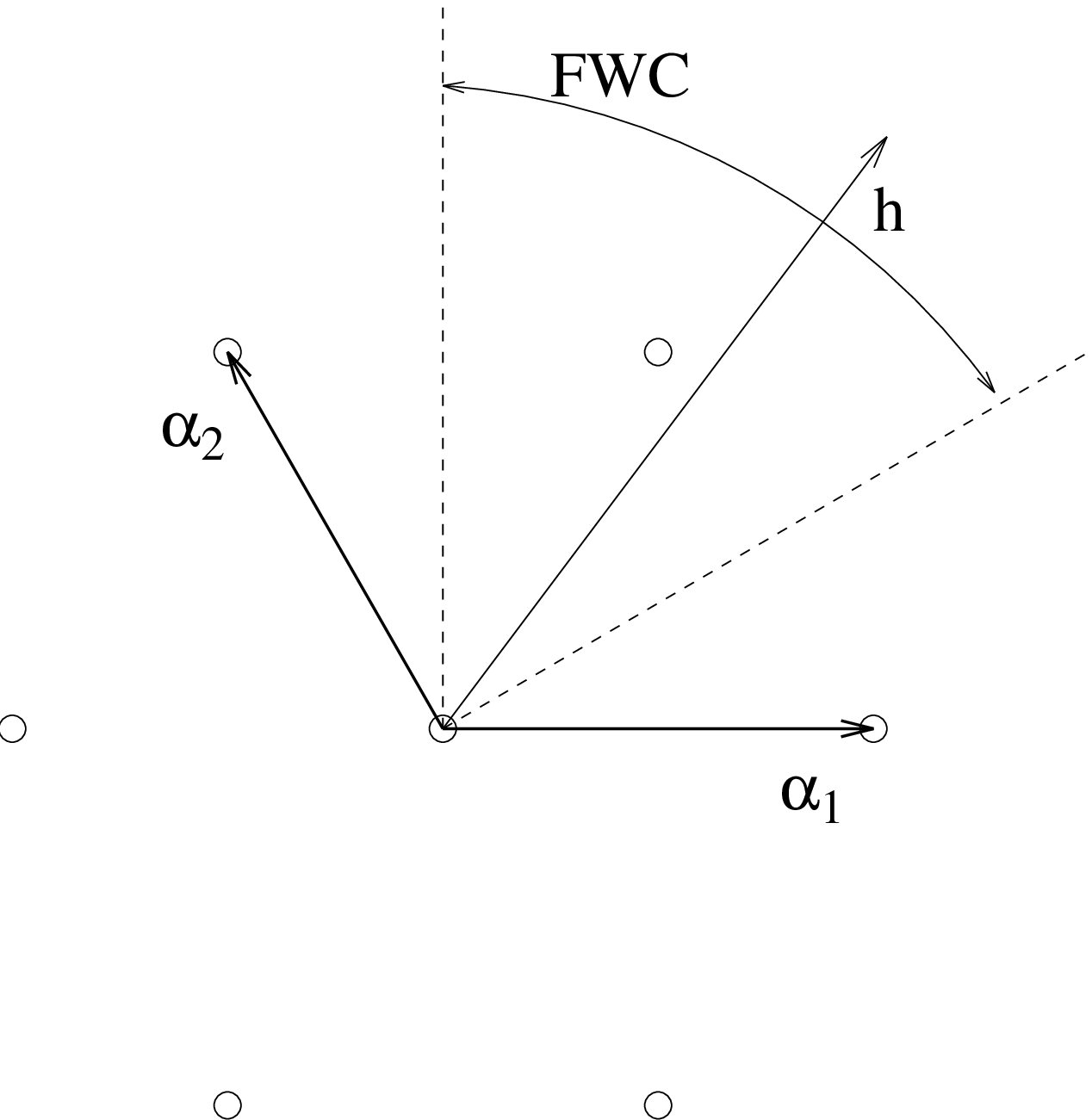}
\vskip.2cm
\centerline{Figure 1. The \fwc\ and the simple roots for $SU(3)$.}
\vskip.3cm
 
Starting with the adjoint \rep, it may be verified that when the magnetic
coroot vector is expressed as a linear combination of the simple coroots
(with the normalization (\corootnormalization)) as
$$
k=k_1\a_1^{\!\vee}+k_2\a_2^{\!\vee}+\ldots+k_r\a_r^{\!\vee}
	\komma\Eqn\kintermsofcoroots
$$
the index for the Dirac operator is
$$
\index\Dslash_{\hbox{\xrm adj}}=2\,(k_1+k_2+\ldots+k_r)\Eqn\adjointindex
$$
for any semisimple Lie group (and maximal symmetry breaking).

In order to translate this result into the complex dimension of a moduli space
at magnetic charge $k$, some care has to be taken ---
it is only true provided that some selfdual configuration 
with the corresponding
asymptotic behaviour of the Higgs field actually exists (or anti-selfdual,
so that the dimension is minus the index). The result indicates
that the real dimension of a moduli space for $k$ a simple coroot is $4$.
This can be verified --- such selfdual solutions exist, and are described by
embeddings of the `t Hooft--Polyakov [\Hooft,\Polyakov] $SO(3)$ monopole.
We denote these simple monopoles.
According to the interpretation of E. Weinberg [\Weinberg], any 
multi-monopole at a $k$ given by (\kintermsofcoroots) with only positive
coefficients $k_i$ can in an asymptotic region be approximated by a 
superposition of well separated simple monopoles, and analogously for
anti-monopoles. This agrees with the linearity of the index in $k$. 
A magnetic coroot formed as (\kintermsofcoroots) with
both positive and negative coefficients would asymptotically correspond 
to a field configuration that is approximately selfdual in some regions
and anti-selfdual in others. Such a configuration can not be static,
since the magnetic and Higgs forces between a monopole and an anti-monopole
do not cancel. Either such configurations do not exist, or they are simply
inaccessible to us at our present understanding. This is of course
a problem already with gauge group $SU(2)$, but there it does not manifest
itself in terms of allowed and disallowed sectors in the coroot lattice,
as it does for higher rank gauge groups, merely as a lack of understanding
of the interaction between monopoles and anti-monopoles.
If one doubts the above argument, it is illuminating to consider the
points in the coroot lattice where the index (\adjointindex) vanishes.
Since the dimensionality of a moduli space can not be zero (translations
are always moduli), it becomes clear that no static BPS configurations
with these magnetic charges can exist. The allowed sectors for magnetic
charges in an $SU(3)$ theory are shown in figure 2, where unfilled roots
indicate forbidden magnetic charges.

Another \rep\ of special interest is the fundamental \rep\ of $SU(N_c)$.
Beginning with $SU(3)$, and ordering the (co)roots by 
$h\cdot\a_1>h\cdot\a_2$, the index becomes $\index\Dslash_{3(SU(3))}\=k_1$.
This is the complex dimension of the fiber of the index bundle of
zero-modes in the fundamental \rep\ for allowed positive magnetic charges.
We note that when the Higgs field aligns with the root $\a_1\+\a_2$ in the
middle of the \fwc, a quark and an antiquark become massless, and
the index formula of Callias may give the wrong result. In fact, when
$h$ crosses this line, the zero-mode at $k\is\a_1$ disappears and a
new zero-mode instead appears at $k\is\a_2$. The index formula gives
a result in between, which clearly is nonsense. The Dirac operator is
not Fredholm in the fundamental \rep\ in this case. It is possible, though,
to follow the asymptotic behaviour of the solutions to the Dirac equation.
For generic $h$ the normalizable solutions decay exponentially with
the radius, while for a degenerate case as this one there is a power law
behaviour. One may check that these solutions have a leading term
proportional to $r^{-1/2}$, so they are not $L^2$. For this special
direction of the Higgs field there are thus no zero-modes in the fundamental
\rep. A similar situation occurs at the boundary of the \fwc, where
the symmetry breaking pattern changes to $H\=SU(2)\cross U(1)$ as
some vector bosons become massless. We do not consider this nonmaximal
symmetry breaking in this paper.

The indices in the fundamental representations of other $SU(N_c)$ groups
behave in a similar way. We can illustrate by looking at $SU(4)$, where
we have the simple roots $\a_{1,2,3}$ with $\a_1^2\is\a_2^2\is\a_3^2\is2$,
$\a_1\sdot\a_2\is\a_2\sdot\a_3\is-1$, $\a_1\sdot\a_3\is0$. 
In the interior of the \fwc\ the symmetry breaking pattern is the maximal
one, $SU(4)\ra U(1)\cross U(1)\cross U(1)$. At the three planes forming
the boundary, $SU(4)$ is broken to $SU(2)\cross U(1)\cross U(1)$, and
where the planes intersect to $SU(2)\cross SU(2)\cross U(1)$ (one line)
or $SU(3)\cross U(1)$ (two lines).
The weights
in the \rep\ 4, specified by their scalar products with the simple
roots, are $\l_{(1,0,0)}$, $\l_{(-1,1,0)}$, $\l_{(0,-1,1)}$ 
and $\l_{(0,0,-1)}$. The \fwc\ divides in two parts, related by the $\Z_2$
of outer automorphisms, and we choose to stay in the region where
$h\sdot\a_1>h\sdot\a_3$. On the boundary there are massless quarks. This
also happens when $h\sdot\l_{(-1,1,0)}\=0$. This plane divides the half \fwc\
in two parts. In the region where $h\sdot\l_{(-1,1,0)}>0$ the index is
$\index\Dslash_{4(SU(4))}\is k_2$ and when $h\sdot\l_{(-1,1,0)}<0$ it is 
$\index\Dslash_{4(SU(4))}\is k_1$. Similar statements hold for higher $SU(N_c)$
groups. The index in the fundamental \rep\ depends only on one of
the simple magnetic charges. 

The \rep\ 6 of $SU(3)$ is interesting because it is contained in
one of the perturbatively finite models.
The index is $\index\Dslash_{6(SU(3))}\=3k_1$, 
with the same choice of ordering of
the roots as above.
The number of complex zero-modes for the \rep s $3$, $6$ and $8$ 
of $SU(3)$ in the allowed sector of positive $k$ are shown in figure 2
(negative $k$ are obtained by reflection in the origin).

\epsfxsize=.6\hsize
\vskip.3cm\hskip2cm\epsffile{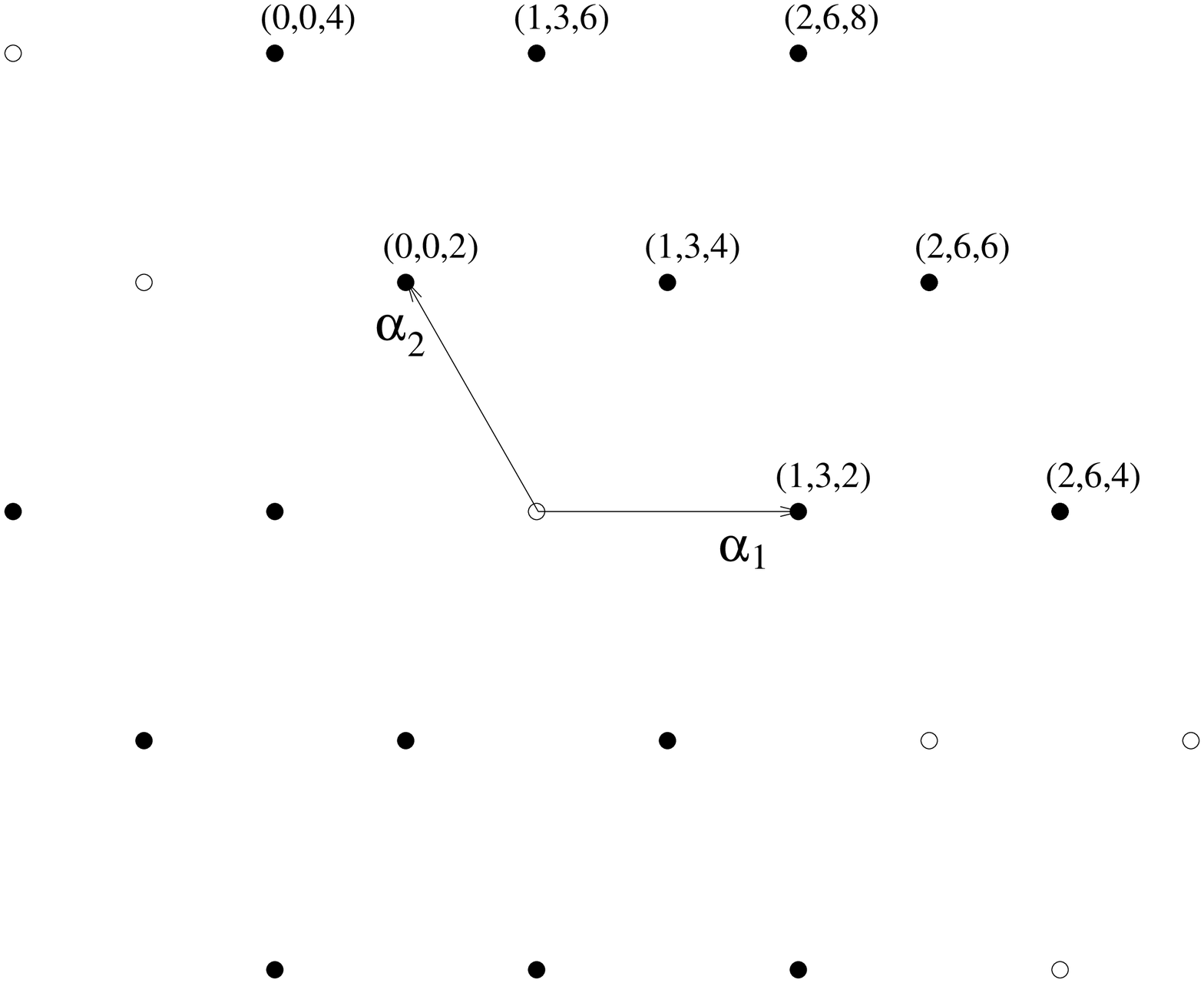}
\vskip.3cm
\centerline{Figure 2. The allowed positive magnetic charges and number of
	zero-modes in 3, 6 and 8 of $SU(3)$.}

\section\lattices{Lattices of Electric and Magnetic Charges --- Duality?}
In this section, we will comment on the possibilities for dual theories
from the viewpoint of electric and magnetic charge lattices.
In the original Goddard--Nuyts--Olive (GNO)
duality conjecture [\GNO], which generalizes
the Montonen--Olive conjecture [\Montonen] of $SO(3)$ 
to higher rank gauge groups
(both applying to $N\is4$ SYM), it is noted that the magnetic charges
lie on the coroot lattice $\Lcr$, 
which is the root lattice of the ``dual group'',
\ie\ the group where long and short roots are interchanged.
It should be noted, for clarity, that even if the spectrum of electric
charges of the elementary excitations of the theory does not span
the entire weight lattice (of $SU(N_c)$, say), this does not leave
us with more choices for the magnetic charges, as one might suspect from
the generalized Dirac condition (\diraccondition). Indeed, the magnetic
charges still are constrained to the coroot lattice, as seen from
equation (\chargevector), disregarding of the matter content of the theory.
So, in the case of $N\is4$ SYM with gauge group $SU(N_c)$, where all
fields come in the adjoint \rep, \ie\ only in one of the $N_c$ conjugacy
classes of the weight lattice, so the actual gauge group is
$SU(N_c)/\Z_{N_c}$, the magnetic charges still lie on the coroot lattice
of $SU(N_c)$, \ie\ on the weight lattice of $SU(N_c)/\Z_{N_c}$.
The GNO conjecture states that an $N\is4$ SYM theory is $\Z_2$ dual
to the $N\is4$ SYM theory with the dual gauge group. The validity
of the conjecture has been partially vindicated by actual calculation
of parts of the dyon spectra [\Sen,\Porrati,\GauntlettIV,\Lee].

When we consider $N\is2$ models with a matter content that makes
the theory perturbatively finite, the GNO interpretation of the
coroot lattice must be revised. For example in the $SU(2)$ theory with
four fundamental hypermultiplets, the coroot lattice of $SU(2)$ is 
reinterpreted as the weight lattice of $SU(2)$ instead of the root lattice.
This simply amounts to a rescaling by a factor $2$. The $\Z_2$ pictures
of the quarks now reside at $k\is\pm\a^{\!\vee}$, where in the $N\is4$ theory
the duals of the massive vector bosons were found. This is of course possible
due to the simple fact that the root and weight lattices of $SU(2)$
are isomorphic up to an overall scale. Some of the dyonic states with
low magnetic charges have been found, and support the duality hypothesis
[\SeibergIII,\Sethi,\GauntlettIII]. 

When we move to more general
gauge groups, the picture is less clear. As a first example, we have
the two perturbatively finite $SU(3)$ theories, one with six hypermultiplets
in the fundamental \rep, the other with one fundamental multiplet
and one in the \rep\ $6$. The elementary excitations now carry electric
charges in all three conjugacy classes of $SU(3)$, so we want also the
magnetic charges to fill out the entire weight lattice of $SU(3)$,
if $\Z_2$ duality is supposed to hold. This reinterpretation of
the coroot lattice is indeed possible, since
the root and weight lattices of $SU(3)$ are isomorphic up to a scale.
It is therefore meaningful to examine the actual spectrum of monopoles
and dyons in these two models in order to find signs for or against
strong--weak coupling duality. As we will see later, the dyon spectra
do not support na\"\i ve selfduality.

In general, already considering the lattices seems to contradict na\"\i ve
duality. The coroot lattice, being the root lattice of the dual group,
is generically not isomorphic to a weight lattice containing \rep s
that allow matter multiplets in other conjugacy classes than the
trivial one. Take $SU(4)$ as an example. The (co)root lattice 
of magnetic charges is the 
fcc lattice, while the weight lattice (dual to the (co)root lattice)
of electric charges is the bcc lattice. With the GNO interpretation of the
coroot lattice, the dual gauge group is $SU(4)/\Z_4$ and there is no
room for matter in nontrivial conjugacy classes. The only possible
matter content is in the adjoint \rep, yielding the $N\is4$ theory.
One might look for a dyon spectrum that only contains states on some
sublattice of the coroot lattice, isomorphic to the relevant part of the
weight lattice
[\Barbon]. Such sublattices exist, but as we will show explicitly
(with $SU(4)$ as an example), the dyon
spectrum is not confined to such a sublattice. Again, we recognize
no signs of selfduality in the dyon spectrum.

Another point, already touched upon in section \indexcalculations,
is that even if the isomorphism between the root and weight lattices
for $SU(3)$ is used as above, or if one tries to pick out a sublattice
isomorphic to (part of) the weight lattice, one is immediately led
to considering states in ``forbidden sectors'', asymptotically
consisting of superpositions of monopoles and anti-monopoles. Such
configurations are not included in the present treatment. Whether this
is a fundamental impossibility or an incompleteness of the semi-classical
procedure is not clear to us (there might exist non-static configurations
that are possible to interpret as bound states of monopoles and
anti-monopoles, although it is unclear to us how such states could
saturate a \Bog\ bound).     

\section\effectiveaction{The Effective Action --- Quantization}
The procedure we follow in order to find the soliton spectrum of the full
quantum field theory is to make a low energy approximation of the theory
in a BPS background. Due to the mass gap, corresponding to the Fredholm
property of section \indexcalculations, the number of degrees of freedom
in this approximation becomes finite. The field configuration moves
adiabatically on the moduli space, and the behaviour of the model is that
of a supersymmetric quantum mechanical model with the moduli space as
target space. The number of supersymmetries is half the number of 
supersymmetries in the original field theory. The reason why this low
energy approximation gives reasonable information about the spectrum of
the full theory is that if we find BPS saturated states at low energy, these
will come in a short multiplet of the $N\is2$ supersymmetry algebra, and
will necessarily continue to do so at any scale [\Olive]. If the theory is
finite, the mass formula of the adiabatic approximation will be exact, while,
if the theory is renormalizable, it is renormalized.

In order to find the supersymmetric quantum mechanical model corresponding
to the actual theory we are interested in, we only keep the zero-modes
of sections \zeromodes\ and \indexcalculations\ as dynamical variables.
Concretely, we derive the low energy action by solving for all fields to
order $n\is1$ as in (\nisonesolution), plug the solutions back into 
the field theory action (\fieldtheoryaction), and keep terms of order $n\is2$.
We then integrate over three-space, using the expressions for metrics,
connections and curvatures of section \zeromodes.
The resulting lagrangian was calculated in [\Cederwall], and reads
$$
{\cal L}=-2\pi h\sdot k+\half g_{mn}\dot X^m\dot X^n+\half g_{mn}\l^mD_t\l^n+
	\half\p^\a D_t\p^\a-{1\/4}F_{mn\a\b}\l^m\l^n\p^\a\p^\b\komma
	\Eqn\QMwithF
$$
Here, we have denoted the fermionic variables, in sections of appropriate
bundles over moduli space, with the same letters that were used in the 
field theory action. The covariant derivatives used on the fermions are defined
as $D_t\l^m\is\dot\l^m\+{\G^m}_{np}\dot X^n\l^p$ and
$D_t\p^\a\is\dot\p^\a\+\w^{\a\b}_m\dot X^m\p^\b$.
If one has $N\is4$ supersymmetry, also the $\p$'s come in the tangent
bundle, and the field strength $F$ is the riemannian curvature.
The lagrangian (\QMwithF) has ``$N\is\half\cross4$'' supersymmetry, meaning
that there are four real supersymmetry generators. They take the form
$$
Q(a)=\l^m{{J^{(a)}}_{m}}^{\lower2pt\hbox{$\scriptstyle n$}}V_n\komma\eqn
$$ 
where $V_m$ is the velocity $g_{mn}\dot X^n$. It is essential, and a necessary
consequence of the existence of these supersymmetries, that $F$ is selfdual,
\ie\ a $(1,1)$-form with respect to all three complex structures. 

When quantizing the supersymmetric quantum mechanical system given by
(\QMwithF), we look for ``ground states'', \ie\ states that continue to
saturate a \Bog\ bound. These are zero energy states for the system given
by the lagrangian without the first term, at least when the electric charge
vanishes. The electric charges modify equation (\masschargerelation) to
$$
m^2=(h\sdot e)^2+({2\pi\/g^2}\,h\sdot k)^2\komma\Eqn\newmasschargerelation
$$
where the coupling constant has been reinstated explicitly (this relation
follows from the form of the extension of the $N\!=\!2$ supersymmetry
algebra).
Consider the solutions (\nisonesolution) to the field equations.
We can use them to derive an explicit expression for the electric
charge density:
$$
D_iE_i=\dot X^mD_i\d_mA_i+\a^*\a-\half\b^*\cross\b\punkt\Eqn\chargedensity
$$
Integrating this over three-space gives a ``topological'' electric charge
from the first term, which is the momentum on the $S^1$ of the moduli 
space. Here, the contribution to the charge density is $\dot X^4D_iD_i\Phi$,
and the electric field is proportional to the magnetic field, with the
proportionality constant being the velocity on $S^1$, so that electric 
charges that arise this way are collinear with the magnetic ones.
The second term does not contribute. Using
$\a\is\d_mA\l^m$, where $\l^m$ are real fermionic oscillators, it
gives after integration $\l^m\l^n\!\!\bra \d_mA^*\d_nA\ket\=0$. Using 
$\b\is\r_\a\p^\a$, the last term becomes
$\p^\a\p^\b\!\!\bra\r^*_\a\cross\r_\b\ket$.
For a complex representation, this may contain an element in the Cartan
subalgebra. A straightforward calculation, using the orthogonality
relations for the zero-modes of the fundamental \rep\ of
$SU(3)$ and magnetic charge $\a_1$, shows that it indeed is
$Q\bar\p\p$, where $Q$ is the U(1) charge of the \rep\ $2$ in the decomposition
$3\ra 2_{1/6}\oplus1_{-1/3}$ under 
$SU(3)\ra SU(2)\cross U(1)$,
the SU(2) being defined by $\a_1$ as in the following section.

A comment on the mass--charge relation: When we find a quantum mechanical
state using the low energy action, we can not expect to find the exact
expression for the mass from the corresponding hamiltonian. What we see
is a low energy approximation. For the $S^1$ momenta, it gives the 
first term in the series expansion for low velocity on the circle.
For the ``orthogonal'' charges from the matter fermions, there is no
continuous classical analogue, and these electric charges are not seen
in the low energy hamiltonian. However, we can deduce from the fact that
the states come in short multiplets that they must be BPS saturated.

One has to divide the fermionic variables into creation and annihilation 
operators. Using the K\"ahler property, we can take $\bar\l^{\bar\mu}$ as
creation operators and $\l^\mu$ as annihilation operators, where $\mu$
is a complex index. We then have two equivalent pictures: either the
states are forms with anti-holomorphic indices, or we view
$\l^m$ as gamma matrices as in the quantization of the spinning string,
and the states are Dirac spinors. The equivalence
is easily seen from a \rep\ point of view --- when the full holonomy
$SO(4n)$ is reduced to $SU(2n)$, the two spinor chiralities decompose
into even and odd forms. Zero energy states are harmonic forms, or spinors
satisfying the Dirac equation.

The presence of the $\p$'s means that the forms/spinors have to be
harmonic with respect to the connection $\w$, and also carry antisymmetric
indices coming from the creation operator part of $\p$ (or a spinor index).
In the case of $N\is4$, the fermions together come in the complexified
tangent bundle, so that ground states are any harmonic forms.

The general pattern is that the part of the $\l$'s belonging to the
$\R^3\cross S^1$ part of moduli space generates the appropriate number
of states of a short multiplet of the space-time supersymmetry algebra.
We thus only have to consider the internal space (and only count singlets
under the discrete group that is divided out) in order to find the
number of multiplets. When the dimension of the internal space is four,
one can use the selfduality of the field strength for a vanishing theorem,
completely analogous to the one used in space-time: all the solutions
to the Dirac equation have to come in only one of the spinor chiralities.
This reduces the problem of identifying the ground states to that of
calculating the index of the Dirac operator. For higher-dimensional
moduli spaces, there is a priori no such vanishing theorem, and it seems
like one has to resort to calculating the $L^2$ cohomology, which of course
is a much harder problem, of which little seems to be known. 

\section\bundles{Dyon Spectra for Low Magnetic Charges}
The moduli spaces for the magnetic charge being any simple coroot is
identical to the one-monopole moduli space in the $SU(2)$ theory. Also,
when $k$ is a multiple of a simple coroot, the moduli space is identical
to the corresponding $SU(2)$ moduli space. The new ingredient for higher
rank gauge groups comes when $k$ is a linear combination of different
simple coroots. If $k$ is a linear combination of orthogonal simple
coroots, the moduli space factorizes metrically into the product of
$SU(2)$ moduli spaces. The only nontrivial example that is accessible 
so far is the space at $k\is\a^{\!\vee}\+\b^\vee$, 
where $\a^{\!\vee}\sdot\b^\vee<0$.
As is pointed out in [\GauntlettIV,\Lee], a very general argument tells 
us that the isometry group of the inner moduli space has to be 
$SU(2)\cross U(1)$ (the ``extra'' $U(1)$ isometry is associated with local
conservation of the ``relative'' magnetic charge). The unique regular
hyperK\"ahler manifold with this isometry is Taub--NUT with positive
mass parameter, and global considerations (see Appendix A)
lead to to the moduli space
$$
\M=\R^3\cross{S^1\cross\,\hbox{Taub--NUT}\/\Z_2}\punkt\eqn
$$
Appendix B contains some basic facts about Taub--NUT space.  

We also would like to find explicit expressions for the connections and
curvatures of the index bundles associated with the various matter fermions.
Starting with the fundamental representation of $SU(3)$ as a model example,
we consider the magnetic charge $k\is\a_1$. 

\epsfxsize=.35\hsize
\vfill
\hskip3cm\epsffile{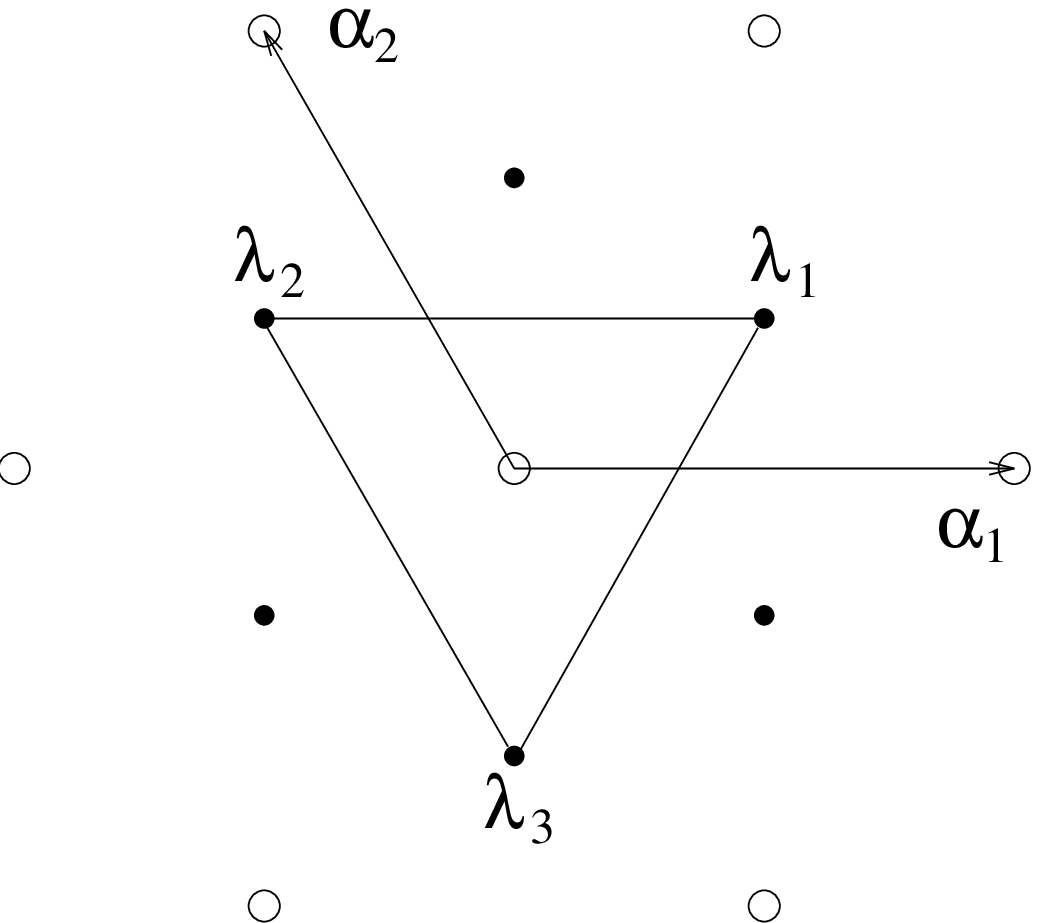}
\vskip.3cm
\centerline{Figure 3. The \rep\ 3 of $SU(3)$.}
\vfill

It is clarifying to calculate the index for the Dirac operator using
a decomposition into $SU(2)\cross U(1)$, where the $SU(2)$ is defined by
the root $\a_1$. The decomposition of the representation $3$ is
$3\ra 2_{1/6}\oplus1_{-1/3}$. Only the $2$ of $SU(2)$ 
(containing the weights $\l_1$ and $\l_2$ of figure 3) has a zero-mode,
so that the zero-modes carry a $U(1)$ electric charge $1/6$. 
The $S^1$ in the moduli
space is generated by gauge transformations with (the SU(2) part of)
the Higgs field as
gauge parameter. Already when this transformation arrives at the
group element $\exp(\pi i\a_1\sdot\,T)$, the nontrivial element in the
center of $SU(2)$, it acts as the identity in the adjoint representation.
In the fundamental \rep\ of $SU(2)$, on the other hand, this element acts
as minus the identity, which means that the index bundle has a $\Z_2$ twist
around the $S^1$ [\MantonII]. This is true also here. If one imposes
single-valuedness of the wave function, this implies that there is a
correlation between the $S^1$
momentum, which is the electric charge in the $\a_1$ direction and
the excitation number of the $\p$'s, carrying electric charge in the
direction orthogonal to $\a_1$. 
The result of these considerations is that the electric charges are constrained
to lie on the weight lattice, which is of course expected.
The electric spectrum at $k\is\a_1$ for the theory with six
fundamental hypermultiplets is indicated in figure 4. The numbers denote
\rep s under the flavour $SU(6)$.

The \rep\ $6$ is treated similarly. It decomposes as 
$6\ra 3_{-1/3}\oplus2_{1/6}\oplus1_{2/3}$. The $3$ carries two 
zero-modes (in the tangent bundle) and the $2$ one. 
This last zero-mode again has a $\Z_2$ twist
around $S^1$. The electric spectrum at $k\is\a_1$ for the model
with one fundamental hypermultiplet and one in the \rep\ $6$ 
is depicted in figure 5, where the number of multiplets at each lattice
point is indicated.

\epsfxsize=.55\hsize
\vskip.3cm
\hskip2cm\epsffile{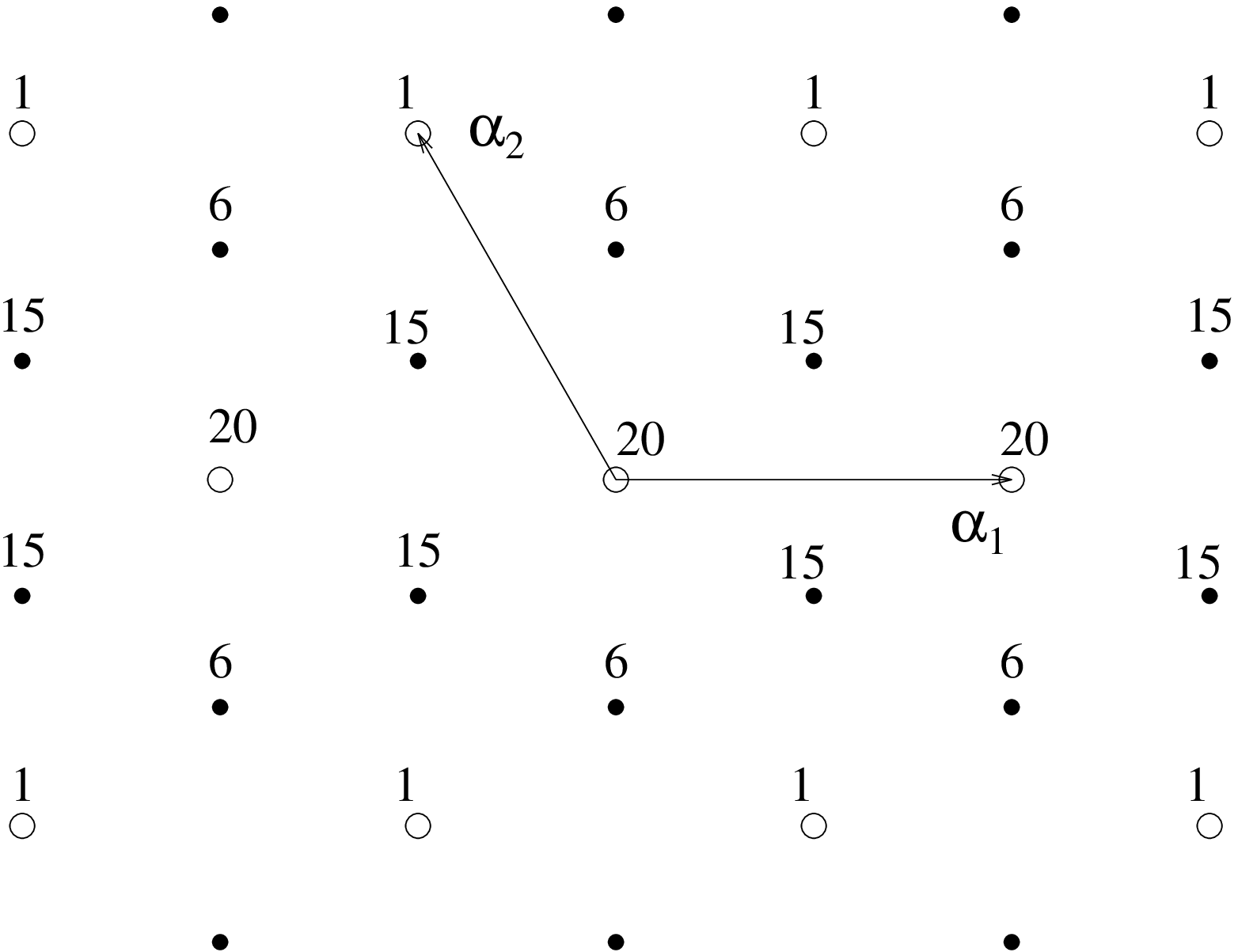}
\vskip.15cm
\centerline{Figure 4. The electric spectrum at $k\is\a_1$ for six multiplets
	in 3 of $SU(3)$.}
\vfill

\epsfxsize=.55\hsize
\vskip.5cm\hskip2cm\epsffile{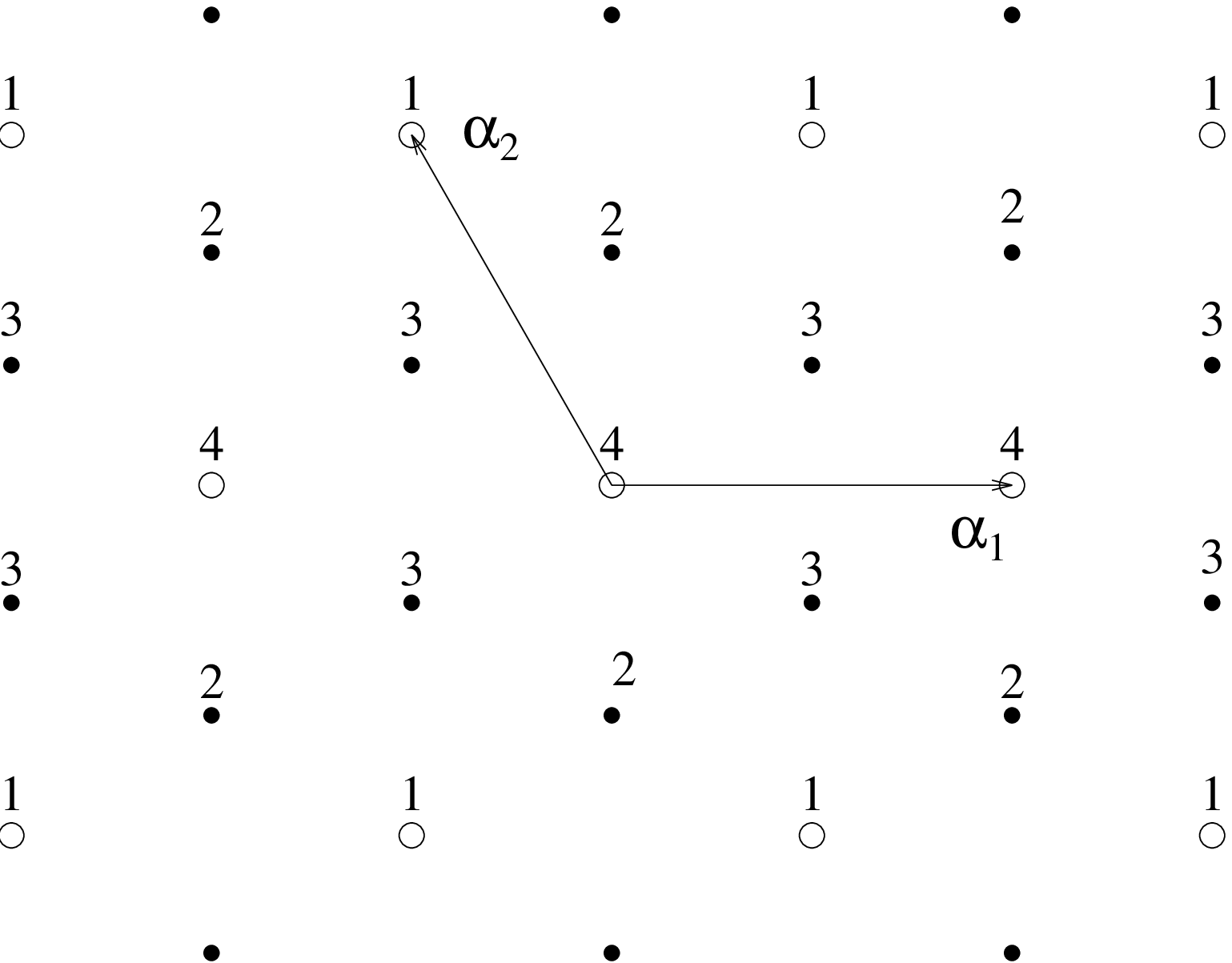}
\vskip.2cm
\centerline{Figure 5. The electric spectrum at $k\is\a_1$ for one 
	multiplet in 3 and one in 6 of $SU(3)$.}
\vfill  
At $k\is\a_2$, there are no matter zero-modes. We just get one multiplet
of states at electric charges that are multiples of $\a_2$. The same 
statement holds
true in the presence of matter in the representation $6$.

At $k\is\a_1\+\a_2$, there is one zero-mode in the fundamental \rep. 
We have to find the connection of the index bundle
over the Taub--NUT space. It has to be a $U(1)$ connection with selfdual
field strength. It is a well known fact that there exists only one
(linearly independent) selfdual harmonic two-form on Taub--NUT space, 
to which the field strength then has to be proportional
(see appendix B). 
We have to determine the normalization factor $c$ in front of the
connection. This can be done by
considering the holonomy in the region of moduli space where the monopoles
are well separated, \ie\ at large $r$. If we move around the circle
generated by ${\*\/\*\p}$, the first time we should get back to the original
configuration is after completing the whole circle. Integrating along
a curve $C_\gamma\,:\,0\leq\p\leq\gamma$ at constant $r$ gives 
$\int_{C_\gamma}\w\=\gamma c{r-M\/r+M}$. 
Thus, the smallest value of $\gamma$ for which
$\exp(i\int_{C_\gamma}\w)\=1$ at infinite radius should be $4\pi$. This gives
$\oint_{C_{4\pi}}\w\is2\pi$, and $c\is\half$.
We then need to find the index of the
Dirac operator for fields of various charges with respect to the $U(1)$
connection. This is completely analogous to the calculation performed
in [\Sethi,\GauntlettIII] for the Atiyah--Hitchin manifold. One can use
the Atiyah--Patodi--Singer index theorem and push the boundary to infinity.
An additional issue here is that if we want to know the spectrum of
the electric charge orthogonal to $\a_1\+\a_2$, we must investigate
how the solutions depend on the coordinate $\p$. Luckily enough, both the
index and the explicit expressions for the mode functions are known 
[\Pope].
If we call the charge of one creation operator for the matter fermions
$1$, the states will come with charges $q$ which are the ``vacuum charge''
plus $n$, where $n$ is the number of creation operators applied.
When the number of matter multiplets is even (we consider self-conjugate
electric spectra) these charges will be integers.
Pope [\Pope] showed that the number of zero-modes of the 
Dirac operator for positive charge
$q$ is $\half q(q\+1)$ and that they depend on the $\p$ coordinate as
$\exp(-\half i\nu\p)$, $\nu\is1\ldots q$, the number of
states at each value of $\nu$ being $\nu$, together with an analogous 
statement for negative $q$.   
Taub--NUT space is simply connected, so the charges are a priori not 
restricted by any quantization rule, and the results in [\Pope] contain
this more general case.
The value of $\nu$ is related to the electric charge in the direction
orthogonal to $\a_1\+\a_2$ by $Q\=\nu/6$ with the normalization for the
$U(1)$ charge used earlier. The $\Z_2$ identification of the moduli
space produces a correlation between $\nu$ and the $S^1$ momentum.
The spectrum of electric charges for $k\is \a_1\+\a_2$ in the $SU(3)$
model with six fundamental hypermultiplets is depicted in figure 6, where
the numbers indicate $SU(6)$ representations. 

\epsfxsize=.55\hsize
\vskip.5cm\hskip2cm\epsffile{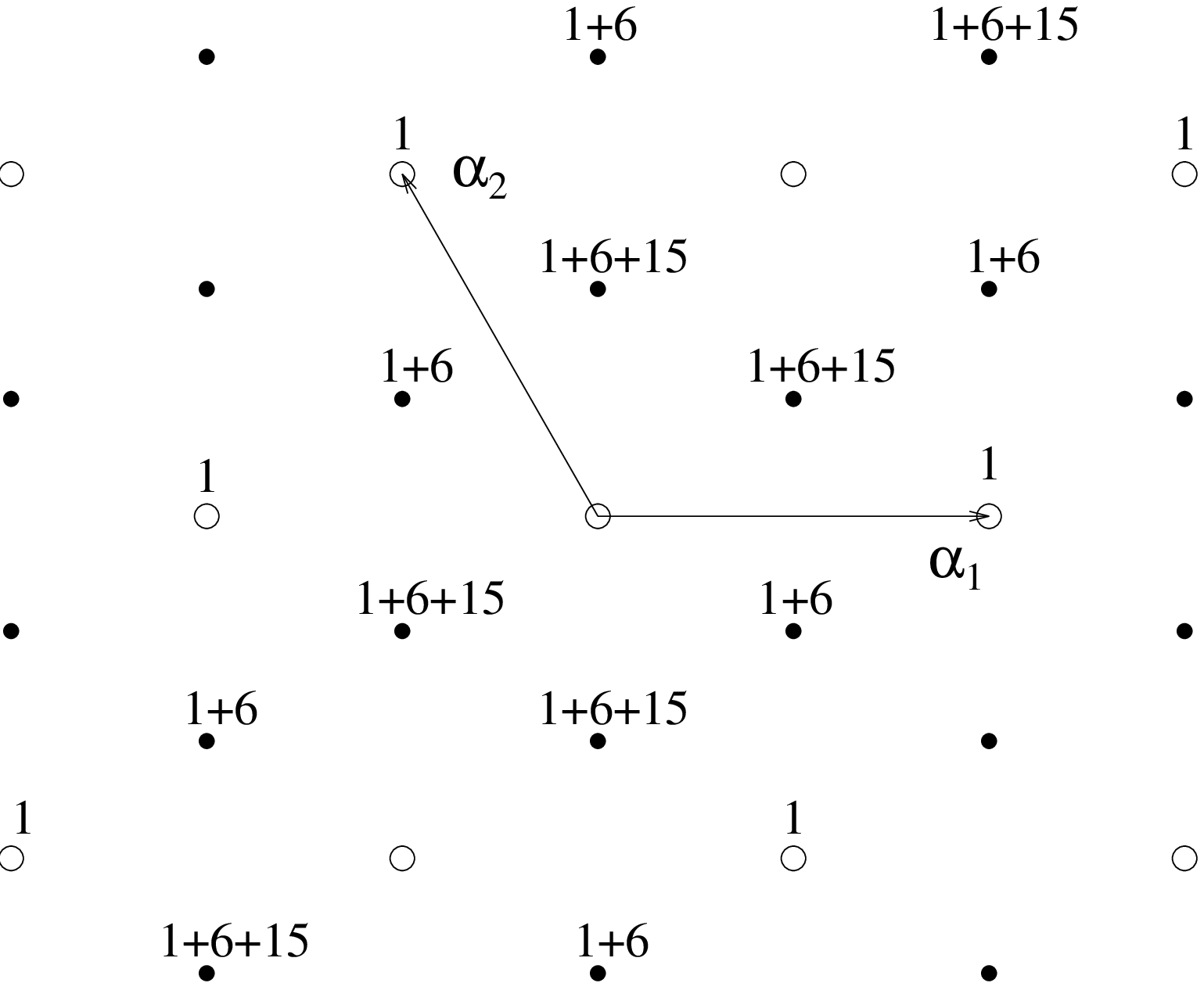}
\vskip.2cm
\centerline{Figure 6. The electric spectrum at 
	$k\is\a_1\+\a_2$ for six multiplets in 3 of $SU(3)$.}
\vskip.3cm
It is probably reasonably straightforward to derive the spectrum at
$k\is \a_1\+\a_2$ also in the presence of matter in the \rep\ $6$.
We have not done this.

The results for the $SU(3)$ theory with six fundamental hypermultiplets
are summarized in figures 4 and 6, together with the electric spectrum
at $k\is\a_2$, which just consists of one multiplet at any integer
multiple of $\a_2$. It is also straightforward to extend the dyon
spectrum to $k\is2\a_1$ and $k\is2\a_2$. We have not been able to 
find the states at the magnetic charges where $\Z_2$ duals of the
vector bosons would be expected to reside. These are either outside
of the allowed sectors or have eight-dimensional inner moduli spaces,
whose metrics are not known. If we examine the electrically uncharged
states at the magnetic charges where the dual quarks were expected,
we find, instead of six multiplet at each lattice point,
twenty, one and zero multiplets at charge $\a_1$, $\a_2$ and $\a_1\+\a_2$
respectively.  

For theories with $N\is4$, the calculations are simpler. As shown in
section \effectiveaction, there are no fermion contributions to the electric
charges, so the electric charge aligns with the magnetic charge.
As already mentioned, ground states correspond to any (normalizable)
harmonic forms on the internal moduli space.
For simple magnetic coroots, the moduli space is $\R^3\cross S^1$, and
there is only one short multiplet for electric charges at integer
multiples of the corresponding root (note that the integer in Dirac's
quantization condition (\diraccondition) is even). For $k$ at twice
a coroot, there is, as demonstrated by Sen [\Sen], a unique selfdual
harmonic two-form on the Atiyah--Hitchin manifold, corresponding to one
multiplet at $e$ being any odd multiple of the root (the selection of
odd multiples comes from the $\Z_2$ divided out in the definition of
the moduli space). At $k$ being the sum of two simple coroots with negative
scalar product, one has, as noted in [\GauntlettIV,\Lee], again the
unique selfdual harmonic two-form mentioned earlier, that now gives
one multiplet at any integer multiple of the corresponding root.
Porrati [\Porrati] has presented convincing evidence for the existence
of {\it all} states predicted by $Sl(2;\Z)$ duality for the $N\is4$ 
$SU(2)$ model. Note that the $Sl(2;\Z)$ duals of the massive vector
bosons in any $N\is4$ theory always lie in the allowed sectors for
the magnetic charges.

When we continue this discussion to higher rank gauge groups, nothing
changes in principle. Part of the above discussion applies to moduli
spaces at simple coroots or sums of two simple coroots for any
gauge group. We have also seen (for the $SU(N_c)$ groups) that the
matter in the fundamental representation behaves very similarly
to what it does in $SU(3)$. This means that we can not hope to find
dyon spectra with magnetic charges confined to a sublattice isomorphic
to the weight lattice. There will always be some states at the simple
coroots, which will not be in such a sublattice.

\section\conclusions{Conclusions and Outlook}
The results of this paper are essentially the following.
In spite of the success of the procedure applied here in finding 
the (low lying)
dyon states predicted by $Sl(2;\Z)$ duality for the $N\is4$ models and
the $N\is2$ $SU(2)$ model with four fundamental hypermultiplets, the
picture we see for higher rank gauge groups and matter content making the
theory perturbatively finite is much less clear. We have for example not
been able to identify the purely magnetically charged states in the
quantum theory with the elementary excitations of some ``dual'' 
finite $N\is2$ theory. There are also sectors of the magnetic charge lattices
that are inaccessible due to our inability of treating systems containing
monopoles and anti-monopoles, and this seems to exclude the treatment
of states needed for duality. This is no problem for the $N\is4$ theories,
since the states needed for duality align with the roots, and are always
found in the allowed sectors, but renders the situation problematic
for $N\is2$ models with gauge groups of rank $r\!\geq\!2$. 

As we see it, there are a couple of possible interpretations
of the results of this paper.
One is that the procedure in some way is incomplete. We saw that some of
the magnetic charges we would need for a duality conjecture lie
in forbidden sectors, that would correspond to superpositions of monopoles
and anti-monopoles, something that is not accessible even in the $SU(2)$
models. We do not know how to describe scattering of monopoles and
anti-monopoles, unless we move to a dual picture. On the other hand,
if such configurations were relevant, they would enter at any magnetic
charge, and they would probably modify the successful calculations
supporting duality for the finite $SU(2)$ model. We find it
unlikely that this could explain any shortcomings. In addition, we have 
seen that the lattice structures have problems that such a modification
hardly could overcome. 

A very drastic explanation of the results would be that the theories
under consideration are not finite --- that there would be instanton
corrections to the $\b$ function, although one has perturbative
finiteness. This sounds very strange and quite unlikely to us, 
but to our knowledge instanton contributions have not been calculated. 
On the other hand, the methods of [\SeibergIV, \SeibergIII] have been applied
to the case of $SU(N_c)$ with fundamental matter 
[\Hanany,\ArgyresI,\ArgyresII,\Minahan], and these results, 
support exact finiteness (although some of the statements are conflicting).
It should be possible to perform at least a one-instanton calculation in order
to verify that these models also are nonperturbatively finite.

A last possibility, which seems most likely, is that there is some 
kind of modified version of
duality that does not include the $\Z_2$ of strong--weak coupling.
A consideration that might give a clue is the following. The duality group
has been conjectured to be not only $Sl(2;\Z)$, but $Sp(r;\Z)$, where
$r$ is the rank of the gauge group. When we examine the dyon spectrum of
the $N\is4$ theories, on the other hand, we only find electric charge
vectors aligned with the magnetic ones (this is a direct consequence of
the properties of monopole configurations at a multiple of a coroot, being
embedded $SU(2)$ monopoles), so that we see only $Sl(2;\Z)$ pictures
of the elementary excitations. When we move to $N\is2$ theories with
higher rank groups, the ``off-diagonal'' part of the $Sp(r;\Z)$ matrices,
\ie\ the one exchanging electric and magnetic charge, consists of a
tensor in $\Lcr\!\otimes\!\Lcr$ and one in $\L_w\!\otimes\!\L_w$.
Of course, in a suitable basis, these just become matrices with integer
entries, but when the basis vectors for the two lattices are not aligned
(which they in general are not, since the lattices are different) such a
basis is not natural, in view of the mass formula (\newmasschargerelation).
This means that in general, and even for $SU(3)$, there is no ``natural''
way of chosing an $Sl(2;\Z)$ subgroup of $Sp(r;\Z)$, where the tensors
mentioned above would become diagonal. A supposed $\Z_2$ duality would in
turn lie in such an $Sl(2;\Z)$ subgroup. One might then speculate in some
kind of ``duality'' for higher rank gauge groups that actually does not
include a $\Z_2$ of electric--magnetic exchange. We find this issue
interesting to pursue. In connection it is also worth mentioning that peculiar
lattice properties of the charges in higher rank gauge groups have been
found earlier. In reference [\ArgyresIII], the existence of simultaneously
massless dyons with nonvanishing $Sp(r;\Z)$ product was demonstrated
(for gauge group $SU(3)$), so that there should exist vacua where elementary
excitations couple both electrically and magnetically to the gauge field.
The evidence points towards a quite rich and interesting structure of 
these theories.
 
In conclusion, the results of this paper, rather than giving definite
answers, raises a number of questions we find it urgent to investigate.

\vskip4\parskip
\noindent\underbar{\it Note added:} After correspondence with the authors of
reference [\Minahan], we realize that for gauge groups of rank 2, 
and only then, there
is a ``natural'' $\Z_2$ transformation, namely where the above mentioned
tensors are the ``epsilon tensors'' 
$\a^{\!\vee}_1\otimes\a^{\!\vee}_2-\a^{\!\vee}_2\otimes\a^{\!\vee}_1$
and $\l_1\otimes\l_2-\l_2\otimes\l_1$. In an orthonormal basis these
become antisymmetric matrices, and do not depend on the choice of
simple coroots or weights. Since they relate two different
vector spaces, they can be thought of as unit matrices. Such a transformation
maps the electric charges of the fundamental representation of $SU(3)$
on coroots, so we do not find support for duality under this $\Z_2$ group.
In [\Minahan], subgroups of $Sp(r;\Z)$ are considered that preserve the
scalar products between roots of $SU(N_c)$ (up to a scale), so that the
transformation of the ``coupling matrix'' only consists of a transformation
of the complex coupling constant. We hope to return to a closer examination
of subgroups that might explain parts of the spectrum we observe (though
it is difficult to conceive how the entire spectra could be generated).
Our attention has also been drawn to reference [\Aharony], where
some of the arguments and results are very close to ours. 
\vskip4\parskip
\noindent\underbar{\it Acknowledgements:} The authors would like to thank
Gabriele~Ferretti, Jerome~Gauntlett, 
Alexander~von~Gussich, Lars Brink, Aleksandar Mikovi\'c, Bengt~Nilsson, 
Per~Salomonson, Kostas~Sfetsos, 
Bo~Sundborg, Per~Sundell and Edward Witten for valuable discussions, 
comments and/or communication of results. We also thank Joseph Minahan
and Dennis Nemeschansky for drawing our attention to reference [\Minahan]
and Ofer Aharony and Shimon Yankielowicz for pointing out the relevance
of reference [\Aharony].
\vfill\eject
\refout
\vfill\eject

\appendix{Topology of the Moduli Space at $k\is\a_1\+\a_2$}
As we have seen, the moduli space at $k\is\a_1\+\a_2$ is actually completely
determined just by considering its isometries together with the
hyperK\"ahler property. In this appendix, we will use the correspondence
between moduli spaces and spaces of rational holomorphic maps to
get direct information about the topology of this space, and support
the indirect arguments. This procedure could in principle be continued
along the lines of [\AtiyahI] to obtain also the metric. 

For $SU(2)$ monopoles, there is an isomorphism between the moduli space
at charge $k$ and the space of rational holomorphic maps $S^2\ra S^2$ ,
due to Donaldson [\Donaldson]. This result was extended to more general
groups by Hurtubise [\Hurtubise], where the case of 
maximal breaking was considered,
and the moduli spaces shown to be isomorphic to spaces of rational 
holomorphic maps from $S^2$ to $G/H$ (``the broken gauge group'').
The target space of the holomorphic map is a ``flag manifold'',
\ie\ a space of nested vector subspaces 
$\C\subset\C^2\subset\ldots\subset\C^N$. This makes it quite straightforward
to write down explicit coordinates for these manifolds as $\C P^1$ bundles
over $\C P^2$ bundles over $\ldots$ over $\C P^{N\-1}$.

We will examine the case of $SU(3)/(U(1)\cross U(1))$, \ie\ the
manifold of complex lines in a complex plane in $\C^{\,3}$.
This clearly implies an $S^2$ bundle over $\C P^2$.
Explicit parametrization of the plane and the line, and some minor 
redefinition in order to make things as symmetric as possible, gives
the coordinates $(x_i,y_i;\z_i)$ in patch $i$, $i\is1,2,3$,  with the overlaps 
$$
(x_{i+1},y_{i+1};\z_{i+1})=(y_i^{-1},x_i^{\phantom{-1}}\hskip-7pt y_i^{-1};
	-\a x_i-\a^{-1}y_i^{\phantom{-1}}\hskip-7pt \z_i^{-1})
\komma\aEqn\overlap
$$
where $\a\is e^{2\pi i\/3}$ and $3\+1$ is understood as $1$.
Here, the $\z$ coordinates are fiber coordinates for $S^2$. We have not cared
to write two separate patches for the fiber, since the one-point 
compactification of $\C$ is trivial. The coordinates for the base manifold
are the standard ones on $\C P^2$. 
If one instead considers the flag ``turned inside out'', \ie\ consider
the complementary (normal) vector subspaces, one is led to an alternative
fibration, given by the coordinate transformations
$$
(\tilde x_i,\tilde y_i;\tilde\z_i)=(\z_{i+1}^{-1},\z_{i+2}^{\phantom{-1}};
	y_{i+1}^{\phantom{-1}})
	\punkt\aEqn\insideout
$$
These coordinates have identical overlap relations as the original ones.
The transformation corresponds to the action of the nontrivial 
element in the $\Z_2$
of outer automorphisms of $SU(3)$.

We now want to examine some simple rational holomorphic maps from $S^2$
to this manifold. These maps should be ``based''. We choose the base point
condition $(x,y;\z)(\infty)\=(1,1;1)$, which is the same in all patches.
It is easy to find a basis for the second homotopy. The fiber $S^2$ 
of course has second homotopy $\Z$, and so has the base manifold $\C P^2$,
being $S^5/U(1)$. The holomorphic maps corresponding to one winding on the
fiber, \ie\ one of the simple magnetic charges, say $\a_1$, 
are easily written down:
$$
\left(\matrix{x_1&y_1&\z_1\cr x_2&y_2&\z_2\cr x_3&y_3&\z_3\cr}\right)(z)
=\left(\matrix{1&1&{z+A\/z+B}\cr 1&1&{z+B\/z+C}\cr 1&1&{z+C\/z+A}\cr}\right)
\komma\aEqn\onezeromap
$$  
where $A\+\a B\+\a^2C\=0$. The easiest way of finding the maps corresponding
to the other simple root $\a_2$ is to apply the coordinate transformation
(\insideout) to the right hand side of (\onezeromap) to obtain
$$
\left(\matrix{x_1&y_1&\z_1\cr x_2&y_2&\z_2\cr x_3&y_3&\z_3\cr}\right)(z)
=\left(\matrix{{z+A\/z+C}&{z+B\/z+C}&1\cr
		{z+C\/z+B}&{z+A\/z+B}&1\cr
		{z+B\/z+A}&{z+C\/z+A}&1\cr}\right)\punkt\aeqn
$$
The corresponding monopoles are the embedded 't Hooft--Polyakov solutions,
and it is easy to deduce that the topology of these moduli spaces is
$\C\cross\C^*\cong\R^3\cross S^1$. A more interesting case is the magnetic
charge $\a_1\+\a_2$. This map winds once around each of the primitive cycles.
We write down the most general ansatz possible, and then derive constraints
on the parameters that enter:
$$
\left(\matrix{x_1&y_1&\z_1\cr x_2&y_2&\z_2\cr x_3&y_3&\z_3\cr}\right)(z)
=\left(\matrix{{z+A\/z+C}&{z+B\/z+C}&{z+D\/z+E}\cr
		{z+C\/z+B}&{z+A\/z+B}&{z+F\/z+D}\cr
		{z+B\/z+A}&{z+C\/z+A}&{z+E\/z+F}\cr}\right)\punkt\aeqn
$$
The outer automorphisms act as $(A,B,C)\leftrightarrow(D,E,F)$.
Using the overlap functions we arrive at the constraints between the
six complex parameters:
$$
\eqalign{&A+D+\a(B+E)+\a^2(C+F)=0\komma\cr
	&AD+\a BE+\a^2CF=0\komma\cr}\aeqn
$$
so that we arrive at the counting of section \indexcalculations\
for the dimension of this moduli space --- it has real dimension eight.

When we investigate the topology, it is useful to consider holomorphic
vector fields on the flag manifold. Some of these will generate holomorphic
isometries on the moduli space. The regular vector fields we consider
take the same form in all three patches (they are the only ones with this
property):
$$
\eqalign{&\V^{(1)}=(1-xy){\*\/\*x}+(x-y^2){\*\/\*y}
		-(\a+y\z+\a^{-1}x\z^2){\*\/\*\z}\komma\cr
	  &\V^{(2)}=(y-x^2){\*\/\*x}+(1-xy){\*\/\*y}
		+(\a y+x\z+\a^{-1}\z^2){\*\/\*\z}\punkt\cr}\aeqn
$$
 There are also the translations on $S^2$, inducing the vector field
$\V^{(3)}\=x'(z){\*\/\*x}+y'(z){\*\/\*y}+\z'(z){\*\/\*\z}$. 
All of these transformations commute.
The transformations induce transformations of the parameters $A,\ldots,F$.
These are better expressed in a basis where the vector fields act diagonally,
$$
\matrix{a=A+B+C\komma\hfill&d=D+E+F\komma\hfill\cr
	b=A+\a B+\a^2C\komma\hfill&e=D+\a E+\a^2F\komma\hfill\cr
	c=A+\a^2B+\a C\komma\phantom{XXX}\hfill&f=D+\a^2E+\a F\punkt\hfill\cr}
		\aeqn
$$
Then $\V^{(3)}$ only acts on $a$ and $d$ as translation, while, if we
denote the induced action of ${i\/\sqrt{3}}(\a\V^{(1)}\-\a^{-1}\V^{(2)})$ 
by $\d_+$ and that of ${1\/3}(\a\V^{(1)}\+\a^{-1}\V^{(2)})$ by $\d_-$, the
action on the moduli parameters is
$$
\eqalign{
&\matrix{\d_+b=2b\komma\phantom{0XX}\hfill&\d_+e=2e\komma\hfill\cr
	\d_+c=c\komma\hfill&\d_+f=f\komma\hfill\cr}\cr&\cr
&\matrix{\d_-b=0\komma\phantom{2bXX}\hfill&\d_-e=0\komma\hfill\cr
	\d_-c=c\komma\hfill&\d_-f=-f\komma\hfill\cr}\cr}\aeqn
$$
while $a$ and $d$ are inert. The constraints are
$$
\eqalign{
&b+e=0\komma\cr
&ae+bd+cf=0\komma\cr}\aeqn
$$
and they are preserved by all the transformations. The transformation $\d_+$
generates the $\C^*$ that together with the $\C$ of $\V^{(3)}$ forms
$\R^3\cross S^1$. The imaginary part of $\d_-$ is a $U(1)$ isometry.
We can chose a location $\theta$ on the $S^1$ by a finite action 
$\exp(i\theta\Im\d_+)$ on some given base point. The parameters $c$ and $f$
are coordinates for the ``inner part'' of the moduli space. By considering
the action of this translation on the total moduli space, we conclude
that the topology is 
$$
\M\cong\R^3\cross{S^1\cross\R^4\/\Z_2}\punkt\aeqn
$$ 
The ``inner''
or ``relative'' moduli space is topologically $\R^4$. This is the topology
of Taub--NUT space with positive mass parameter.

\appendix{Taub--NUT Space --- Metric and Connections}
This appendix contains a short summary about Taub--NUT space
(see \eg\ reference [\Gibbons] for more detailed discussions).
Taub--NUT space is a member of a very restricted family of four-dimensional
regular hyperK\"ahler manifolds with $SO(3)$ isometry [\Gibbons], 
that also includes
the Atiyah--Hitchin manifold (contained in the moduli space
for magnetic charge twice a simple coroot), and the Eguchi--Hansson manifold.
The properties obtained from simple physical considerations, that
the metric asymptotically approaches $\R^3\cross S^1$ and that the isometry
is $SU(2)\cross U(1)$, singles out Taub--NUT as the internal moduli space
for magnetic charges that are the sum of two simple coroots with
negative scalar product.

The metric may be written 
$$
g={r+M\/r-M}\,dr\otimes dr+(r^2-M^2)(\s_1\otimes\s_1+\s_2\otimes\s_2)
	+4M^2\,{r-M\/r+M}\,\s_3\otimes\s_3\komma
	\aEqn\taubnutmetric
$$  
where the ranges of the coordinates are $M\leq r$, $0\leq\theta\leq\pi$,
$0\leq\phi<2\pi$ and $0\leq\p<4\pi$, and the $\s_i$ are left-invariant
one-forms on $S^3\cong SU(2)$:
$$
\eqalign{&\s_1=\cos\p d\theta+\sin\p\sin\theta d\phi\komma\cr
	&\s_2=-\sin\p d\theta+\cos\p\sin\theta d\phi\komma\cr
	&\s_3=d\p+\cos\theta d\phi\komma\cr}
		\aEqn\sthreeoneforms
$$
with the dual vector fields $v_i$, $v_i(\s_j)\=\d_{ij}$:
$$
\eqalign{&v_1=\cos\p{\*\/\*\theta}+{\sin\p\/\sin\theta}{\*\/\*\phi}
		-\cot\theta\sin\p{\*\/\*\p}\komma\cr
	&v_2=-\sin\p{\*\/\*\theta}+{\cos\p\/\sin\theta}{\*\/\*\phi}
		-\cot\theta\cos\p{\*\/\*\p}\komma\cr
	&v_3={\*\/\*\p}\punkt\cr}\aeqn
$$
If we write the vierbein one-forms as $e_r\=fdr$, $e_i\=c_i\s_i$, the
functions $f$, $c_i$ satisfy 
(prime denotes differentiation with respect to $r$)
$$
{c_1'\/f}={c_1^2-(c_2-c_3)^2\/2c_2c_3}\qquad\hbox{and cyclic.}\aeqn
$$
This equation enables us to calculate the curvature quite easily:
$$
\eqalign{
&R_{0i}=\half\e_{ijk}R_{jk}\cr
&R_{01}=-k_1'dr\wedge\s_1+(-k_1+k_2+k_3-2k_2k_3)\s_2\wedge\s_3
	\qquad\hbox{and cyclic,}\cr}\aeqn 
$$
where $k_i\={c_i'\/f}$. The first equation states that $R$ is selfdual.
The curvature may then be used in the calculation of the index of
the Dirac operator [\Pope], using the Atiyah--Patodi--Singer index theorem
[\AtiyahIII] and pushing the boundary to infinite radius.

When we consider matter zero-modes, we will need a U(1) connection on 
Taub--NUT space with selfdual field strength. There is exactly one
selfdual harmonic two-form (up to normalization).
It is
$$
F=c\,\left({2M\/(r+M)^2}dr\wedge\s_3-{r-M\/r+M}\s_1\wedge\s_2\right)
	\punkt\Eqn\taubnutfeildstrength
$$
The corresponding potential is 
$$
\w=c\,{r-M\/r+M}\,\s_3\punkt\Eqn\taubnutconnection
$$
The coefficient $c$ is determined by physical considerations.

\end